\documentclass[manuscript,screen]{acmart}
\AtBeginDocument{%
  \providecommand\BibTeX{{%
    \normalfont B\kern-0.5em{\scshape i\kern-0.25em b}\kern-0.8em\TeX}}}
\pdfoutput=1
\usepackage[shortlabels]{enumitem}
\usepackage{url,amsmath}
\usepackage{algorithm}  
\usepackage{algorithmic}
\usepackage{subcaption}
\usepackage{bm} 
\usepackage{multirow} 
\usepackage{soul,color}
\usepackage{xcolor}
\usepackage{diagbox}
\usepackage{bbm}
\usepackage{graphicx}
\usepackage{enumitem,pgfplots}
\usepackage{tikz}

\newcommand{\R}{\ensuremath{\mathbb{R}}}

\usepackage{array}
\newcolumntype{x}[1]{>{\centering\arraybackslash\hspace{0pt}}p{#1}}
\acmJournal{TOPS}
\begin{document}

\date{}

\title{\Large \bf Do You Think You Can Hold Me? The Real Challenge of Problem-Space Evasion Attacks}

\author{Harel Berger}
\authornote{Corresponding author.}
\email{harel.berger@msmail.ariel.ac.il}
\orcid{1234-5678-9012}
\affiliation{
  \institution{Ariel Cyber Innovation Center, Computer Science Department , Ariel University}
  \city{Ariel}
  \state{Israel}
  \country{Israel}
}
\author{Amit Dvir}
\email{amitdv@g.ariel.ac.il}
\orcid{1234-5678-9012}
\affiliation{
  \institution{Ariel Cyber Innovation Center, Computer Science Department, Ariel University
 }
  \city{Ariel}
  \state{Israel}
  \country{Israel}
}
\author{Chen Hajaj}
\email{chenha@ariel.ac.il}
\orcid{1234-5678-9012}
\affiliation{
  \institution{Ariel Cyber Innovation Center, Data Science and Artificial Intelligence Research
Center, Industrial Engineering and Management Department, Ariel University}
  \city{Ariel}
  \state{Israel}
  \country{Israel}
}
\author{Rony Ronen}
\email{ronen@msmail.ariel.ac.il}
\affiliation{
  \institution{Ariel Cyber Innovation Center, Computer Science Department, Ariel University}
  \city{Ariel}
  \state{Israel}
  \country{Israel}
}
\begin{abstract}
Android malware is a spreading disease in the virtual world. Anti-virus and detection systems continuously undergo patches and updates to defend against these threats. Most of the latest approaches in malware detection use Machine Learning (ML). Against the robustifying effort of detection systems, raise the \emph{evasion attacks}, where an adversary changes its targeted samples so that they are misclassified as benign.  
This paper considers two kinds of evasion attacks: feature-space and problem-space.
\emph{Feature-space} attacks consider an adversary who manipulates ML
features to evade the correct classification while minimizing or constraining the total manipulations.
\textit{Problem-space} attacks refer to evasion attacks that change the actual sample.
Specifically, this paper analyzes the gap between these two types in the Android malware domain. 
The gap between the two types of evasion attacks is examined via the retraining process of classifiers using each one of the evasion attack types. The experiments show that the gap between these two types of retrained classifiers is dramatic and may increase to 96\%. Retrained classifiers of feature-space evasion attacks have been found to be either less effective or completely ineffective against problem-space evasion attacks. Additionally, exploration of different problem-space evasion attacks shows that retraining of one problem-space evasion attack may be effective against other problem-space evasion attacks.
\end{abstract}

\maketitle

\section{Introduction}
\label{introduction}
Malicious files are a nuisance for ordinary users and security experts alike. However, the task of generating suitable and simple defense systems against the spreading disease of malicious files around the world is on the experts' shoulders. The main course of action for malware detection in recent years is Machine Learning~\cite{aafer2013droidapiminer, arp2014drebin, onwuzurike2019mamadroid, shabtai2009detection,lin2022secure,berger2022mamadroid20}. Several malware ML-based detection systems implement static analysis of the components of the application~\cite{arp2014drebin,demontis2017yes,li2019android,li2020adversarial}. Other methods explore API sequences or the structure of the commands inside the app~\cite{backes2016reliable,onwuzurike2019mamadroidold,xu2018cdgdroid,zhiwu2019android}.  

Some of the ML-based detection systems were proven by Goodfellow et al.~\cite{goodfellow2014explaining} to be vulnerable to manipulations.
\textit{Adversarial examples}, the formal terminology of these manipulations, are generated by an adversary that changes samples to achieve misclassification~\cite{grosse2017adversarial,kuppa2019black,yuan2019adversarial}. \textit{Evasion attacks} (EA),  refer to adversaries that change malicious instances such that they will be classified as benign. 
EAs can be divided into two types: feature-space attacks and problem-space attacks. Feature-space attacks map the malware sample to a feature vector. Then, add perturbations to the values of the vector. Problem-space attacks change the actual sample. Feature-space evasion attacks utilize ML methods that change feature values to evade correct classification. This approach is general and may suit any domain in which an attacker manipulates a malware sample. Still, feature-space attacks do not consider any realistic constraint embedded in the specific malware domain. For example, deleting a sensor from a malicious Android app, thereby setting the associated feature to 0 (from 1) may damage the functionality of the app and consequently disable the malicious activity of the app. In comparison, problem-space attacks are more concrete and require an expert in the specific domain to generate them. 

Most malware is engineered for a specific operating system and may be based on the hardware and functionalities of the victim (as shown in~\cite{avtech}). Therefore, suitable solutions are devised for each platform and domain separately.
Among other domains, the Android operating system is very attractive to attackers, as it is still the most popular mobile operating system in the world~\cite{mobstat}. As a countermeasure, many works in the security community focus on the correct identification of malware lurking in the Android markets~\cite{arp2014drebin,cai2018towards,chen2016stormdroid,dini2012madam,huynh2017new,onwuzurike2019mamadroid,shabtai2012andromaly,shabtai2014mobile,wu2012droidmat,berger2021crystal}. Other studies enhance existing work and make them more robust. For example, retraining detection systems on evasion attacks~\cite{li2020adversarial,lindorfer2015marvin,yuan2019lightweight,nomura2021auto}. The current work examines the Android OS domain and explores the efficiency of approaches to robust ML detection systems with retraining on different evasion attacks. Compared to previous works, this work explores retraining on both problem-space and feature-space evasion attacks. The identification of gaps between these two types of retraining in this study proves that retraining on different types of evasion attacks creates different improvements in detection rates. Specifically, sufficient improvement was not demonstrated in feature-space attacks. Moreover, in some cases, the initial detection rate decreased in  feature-space attacks.

The contribution of this work is twofold: First, this work demonstrates that retraining a basic classifier on feature-space evasion attacks does not gain sufficient robustness against problem-space evasion attacks.
As explained above, feature-space evasion attacks create an abstraction of reality. As a consequence, retraining a basic classifier on feature-space evasion attacks does not strengthen the classifier against problem-space evasion attacks on a specific domain.
On the other hand, retraining a basic classifier on problem-space evasion attacks in a specific domain (e.g., the Android domain) is efficient for problem-space attacks in the same domain. Second, this work analyzes the effectiveness of different defense methods based on retraining a basic classifier on one problem-space evasion attack against other problem-space evasion attacks. In this work it is proven that some of these defense methods can be generalized. Three detection systems were examined for these analyses: Drebin~\cite{arp2014drebin}, an additional version of Drebin using a DNN~\cite{li2020adversarial}, and MaMaDroid~\cite{onwuzurike2019mamadroid} (the first and the latter detection systems are two well-known Android malware detection systems). 

The remainder of this paper is as follows. First, the ML fundamentals of malware detection are described in Section~\ref{fund}. Then, the aspects of the retraining process are presented in Section~\ref{S:framework}. The background on APK structure and on feature sets explored in this study is presented in Section~\ref{background}. An overview of the experimental design is described in Section~\ref{S:exp}. The results of the experiments of retrained classifiers in both feature-space and problem-space evasion attacks are presented in Sections~\ref{S:validation}-\ref{PSR}. Related work is discussed in Section~\ref{related-work}. The conclusions are presented in Section~\ref{conclusion}.   
\section{Fundamentals of Machine Learning and Malware Detection} 
\label{fund}
This section surveys the fundamentals of malware detection and machine learning. First, the formalities of ML are presented in Section~\ref{learn_pred}. Then, the definitions of evasion attacks and mitigation techniques are described in Sections~\ref{ev_conc}-\ref{def_conc}. 
\subsection{ML Formalities}
\label{learn_pred}
The input for a classification problem using supervised machine learning is a training dataset $\mathsf{D} = \{(x_i,y_i)\}$, where $x_i \in \mathsf{X} \subseteq \R^n$ are feature vectors taken from a feature space $\mathsf{X}$, and $y_i \in \mathsf{L}$ are object labels from space $\mathsf{L}$.
Each sample in $\mathsf{D}$ is assumed to be generated i.i.d. from the same unknown distribution $\mathsf{P}$. The goal of this kind of problem is to identify and train a learning model $h \in \mathsf{H}$ (where $\mathsf{H}$ is the hypothesis/models' space), which has the minimal expected error regarding new samples drawn from the same distribution $\mathsf{P}$.

In malware detection, one of the supervised machine learning applications, the feature vectors are not given. Instead, the initial data set comprises a set of entities, such as PDF files, with their labels. To create the feature vectors, a preprocessing operation of the entities is required. Feature extractors generate the mapping of entities-to-feature vectors. A simple example is the Android version of an APK file, which can be easily obtained from the OS and then translated to a numerical value. Employing a feature extractor for every entity in the dataset, and then associating the entity with its object label, resulting in a dataset $\mathsf{D}$ that fits a standard ML system.

\subsection{Evasion Attacks}
\label{ev_conc}
In an \emph{evasion attack}, a learned ML model $h(x)$ is given. This model returns a label $y = h(x)$ for any random feature vector $x \in \mathsf{X}$. In the case of Android malware detection, the label is malicious or benign and the feature vector is extracted from an APK using a feature extractor.

The adversary starts with an initial entity $e$ (such as a malicious APK). From this entity, the attacker extracts a feature vector $x=\phi(e)$. The next steps depend on the type of evasion attack that the attacker runs. If the attacker runs a problem-space evasion attack, it manipulates the actual entity $e$, thus creating $e'$. This new $e'$ entity, is associated with the correlative feature vector $x' = \phi(e')$. Alternatively, if the attacker runs a feature-space evasion attack, it transforms the feature vector $x$ to $x'$, which abstractly describes a transformed entity $e'$, following $e' = \phi^{-1}(x')$ ($\phi^{-1}$ is the inverse function of $\phi$).
The goals of these two kinds of attacks are twofold: First, the label of $x'$, $h(x')$ should return as inaccurate (the attack succeeds). Second, $e'$ should retain the functionality of $e$. In the case of feature-space attacks, the functionality of $x'$ is abstracted, because running a feature vector $x'$ is not feasible.
A cost function, $c(x,x')$, penalizes great modifications of $x'$. This cost function accounts for various changes that may harm the functionality of the theoretic entity $e'$.

Creating problem-space evasion attacks is not easy~\cite{berger2020evasion,oakland2014,ndss2016,tong2019improving}, as a clear understanding of the malware samples structure is required. Irresponsible transformation of $e$ may harm the functionality of $e'$. However, some significant works have demonstrated a method for this (e.g.,~\cite{chen2018android,berger2021crystal,berger2020evasion,pierazzi2020intriguing}). On the other hand, it is quite easier to run feature-space attacks, because running them only changes numeric values. However, this kind of attack may suffer from
the inverse feature-mapping problem~\cite{biggio2013evasion,Biggio13,huang2011adversarial,maiorca2013looking,maiorca2018towards,quiring2019misleading,pierazzi2020problemspace,bostani2021evadedroid}). The \textit{inverse feature-mapping problem} emphasizes the fact that feature-space attacks utilize an abstraction of reality, using a mapping function $\phi$ of the entity $e$ to the feature vector $x$. Manipulations on $x$ by feature-space attacks may not represent feasible changes in the practical domain of the entity $e$. Therefore, applying the inverse function $\phi^{-1}$ to a manipulated feature vector $x'$ can result in a non-functional $e'$. 
\subsection{Mitigation Techniques}
\label{def_conc}
Evasion attacks are a great threat to malware detection systems. Therefore, multiple research studies have explored the attacks' mitigation techniques~(e.g., \cite{pkdd2013, Bruckner12,kdd2011, jmlr2009, Papernot16, Papernot18, Raghunathan18,Vorobeychik18book,Wong18,tong2017framework,berger2022mamadroid20}).
Among these approaches, there are three that should be noted: (a) game-theoretic reasoning, (b) robust optimization, and (c) iterative adversarial retraining.
Some of these approaches cannot be used on every type of ML detection system, such as robust optimization, because solving these problems requires a special structure, such as a continuous feature space~\cite{pkdd2013, Bruckner12,kdd2011}.
On the other hand, iterative adversarial retraining does not include any assumption about the detection system or the attacker~\cite{li2016,tong2017framework}.
In light of the fact that this study involves problem-space evasion attacks and different types of feature values (binary and real numbers), iterative adversarial retraining is the only defense mechanism that can be applied to each of the learning models.
\section{Aspects of Robustification Attempts}
\label{S:framework}

This work explores two main aspects:
\begin{enumerate}
    \item \emph{Validation of feature-space retraining}: Evaluation of the robustness gained by feature-space evasion attacks retraining in the presence of problem-space evasion attacks.
    \item \emph{Generalizability of evasion attacks}: Examining the generalizability of retraining processes  against evasion attacks.
\end{enumerate}

These aspects are explored using the abstraction of a defender and an attacker\footnote{A similar approach was suggested in previous work, formulated as a Stackelberg game~\cite{kdd2011,nips2014,Vorobeychik18book,tong2019improving}.}. The defender chooses an ML defense, which in this work is a learned model $h(x)$. The attacker reacts to this defense with an attack $O(h;\mathsf{D})$, namely an attack that generates manipulated samples from a given dataset $\mathsf{D}$, and the learned model $h$. A measurement, $u$, is formulated, using the formula $u(h;O(h;\mathsf{D}))$. This formula depicts the accuracy of the classifier $h$ on the manipulated samples produced by the attack, $O(h;\mathsf{D})$. The goal of the defender is to optimize its defense, using the following optimization problem:
\begin{equation}
\label{E:robustML}
\max_{h} u(h;O(h;\mathsf{D})).
\end{equation}

\emph{Iterative adversarial retraining} (in short, iterative retraining) is used to solve the optimization problem (\ref{E:robustML}). In this method, malicious instances are manipulated by $O(h;\mathsf{D})$ and iteratively added to the training data. Then, the classifier is adjusted by retraining with the new training data. Thus, the classifier learns the evasion attack. However, using a large number of iterations may create a tendency toward this attack. In other words, the model may be overfitting to this attack. Consequently, the classification of clean data may be damaged. Therefore, the number of iterations should be chosen in correlation with the effect on the classification of clean data.  The iterative adversarial retraining approach was previously proposed to harden classifiers against evasion attacks~\cite{li2016,icml2016,tong2019improving}.
The variant of iterative retraining that was used in this work follows the following steps:
\begin{enumerate}
\item Initial a classifier, $h$.
\item Execute the \emph{evasion attack} on several malicious samples from the
  training data to generate new feature vectors and test the classifier on them.
\item Add all of the new feature vectors to the training data and retrain the classifier.
\item Stop after a fixed number of iterations, or when no
  new samples can be added.
\end{enumerate}

Following the above, the evaluation of the two aspects of the retraining method is now described.

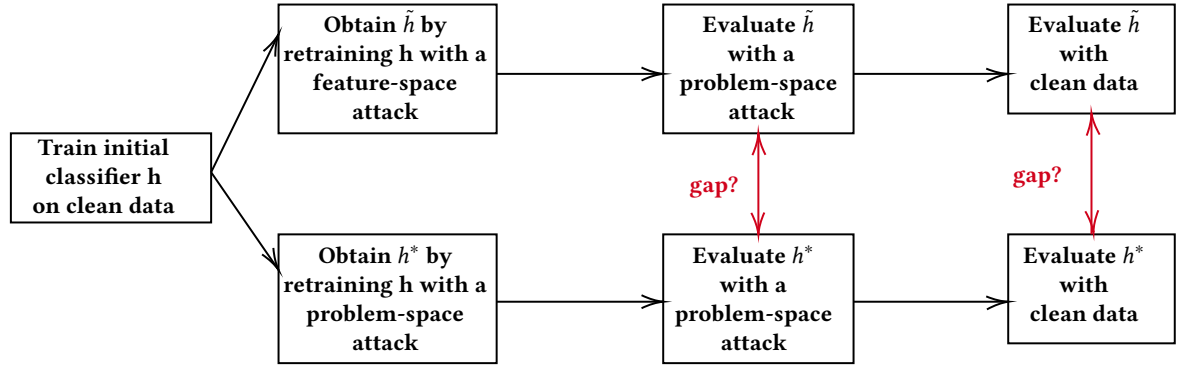
\begin{figure*}[h]

\tikzset{every picture/.style={line width=0.75pt}} 
\begin{tikzpicture}[x=0.75pt,y=0.75pt,yscale=-1,xscale=1]
\draw (519,95) node [anchor=north west][inner sep=0.75pt]   [align=left] {\textcolor[rgb]{0.82,0.01,0.11}{\textbf{gap?}}};
\draw (356,100) node [anchor=north west][inner sep=0.75pt]   [align=left] {\textcolor[rgb]{0.82,0.01,0.11}{\textbf{gap?}}};

\draw    (518,131) -- (602,131) -- (602,186) -- (518,186) -- cycle  ;
\draw (519,135) node [anchor=north west][inner sep=0.75pt]   [align=left] {\begin{minipage}[lt]{53.75pt}\setlength\topsep{0pt}
\begin{center}
\textbf{Evaluate $h^*$}\\\textbf{with }\\\textbf{clean data}
\end{center}

\end{minipage}};
\draw    (344,131) -- (440,131) -- (440,196) -- (344,196) -- cycle  ;
\draw (336,135) node [anchor=north west][inner sep=0.75pt]  [xslant=0] [align=left] {\begin{minipage}[lt]{78.12pt}\setlength\topsep{0pt}
\begin{center}
\textbf{Evaluate $h^*$ }\\\textbf{with a}\\\textbf{problem-space }\\\textbf{attack}
\end{center}

\end{minipage}};
\draw    (150,131) -- (260,131) -- (260,196) -- (150,196) -- cycle  ;
\draw (140,135) node [anchor=north west][inner sep=0.75pt]   [align=left] {\begin{minipage}[lt]{93.4pt}\setlength\topsep{0pt}
\begin{center}
\textbf{Obtain $h^*$ by }\\\textbf{retraining h with a }\\\textbf{problem-space}\\\textbf{attack}
\end{center}

\end{minipage}};
\draw    (518,15) -- (602,15) -- (602,70) -- (518,70) -- cycle  ;
\draw (519,16) node [anchor=north west][inner sep=0.75pt]   [align=left] {\begin{minipage}[lt]{53.75pt}\setlength\topsep{0pt}
\begin{center}
\textbf{Evaluate $\tilde{h}$}\\\textbf{with }\\\textbf{clean data}
\end{center}

\end{minipage}};
\draw    (344,15) -- (440,15) -- (440,80) -- (344,80) -- cycle  ;
\draw (340,16) node [anchor=north west][inner sep=0.75pt]  [color={rgb, 255:red, 208; green, 2; blue, 27 }  ,opacity=1 ,xslant=0] [align=left] {\begin{minipage}[lt]{78.12pt}\setlength\topsep{0pt}
\begin{center}
\textbf{\textcolor[rgb]{0,0,0}{Evaluate $\tilde{h}$ }}\\\textbf{\textcolor[rgb]{0,0,0}{with a}}\\\textbf{\textcolor[rgb]{0,0,0}{problem-space }}\\\textbf{\textcolor[rgb]{0,0,0}{attack}}
\end{center}

\end{minipage}};
\draw    (150,15) -- (260,15) -- (260,80) -- (150,80) -- cycle  ;
\draw (140,16) node [anchor=north west][inner sep=0.75pt]   [align=left] {\begin{minipage}[lt]{93.4pt}\setlength\topsep{0pt}
\begin{center}
\textbf{Obtain $\tilde{h}$ by }\\\textbf{retraining h with a }\\\textbf{feature-space}\\\textbf{attack}
\end{center}

\end{minipage}};
\draw    (15,80) -- (116,80) -- (116,125) -- (15,125) -- cycle  ;
\draw (15,81) node [anchor=north west][inner sep=0.75pt]   [align=left] {\begin{minipage}[lt]{67.91pt}\setlength\topsep{0pt}
\begin{center}
\textbf{Train initial }\\\textbf{classifier h }\\\textbf{on clean data}
\end{center}

\end{minipage}};
\draw [color={rgb, 255:red, 208; green, 2; blue, 27 }  ,draw opacity=1 ]   (560,70) -- (560,131) ;
\draw [shift={(561,130)}, rotate = 265.6] [color={rgb, 255:red, 208; green, 2; blue, 27 }  ,draw opacity=1 ][line width=0.75]    (10.93,-3.29) .. controls (6.95,-1.4) and (3.31,-0.3) .. (0,0) .. controls (3.31,0.3) and (6.95,1.4) .. (10.93,3.29)   ;
\draw [shift={(560,70)}, rotate = 85.6] [color={rgb, 255:red, 208; green, 2; blue, 27 }  ,draw opacity=1 ][line width=0.75]    (10.93,-3.29) .. controls (6.95,-1.4) and (3.31,-0.3) .. (0,0) .. controls (3.31,0.3) and (6.95,1.4) .. (10.93,3.29)   ;
\draw    (440,165) -- (518,165) ;
\draw [shift={(518,165)}, rotate = 176.87] [color={rgb, 255:red, 0; green, 0; blue, 0 }  ][line width=0.75]    (10.93,-3.29) .. controls (6.95,-1.4) and (3.31,-0.3) .. (0,0) .. controls (3.31,0.3) and (6.95,1.4) .. (10.93,3.29)   ;
\draw [color={rgb, 255:red, 208; green, 2; blue, 27 }  ,draw opacity=1 ]   (392,80) -- (392,131) ;
\draw [shift={(392,131)}, rotate = 268.53] [color={rgb, 255:red, 208; green, 2; blue, 27 }  ,draw opacity=1 ][line width=0.75]    (10.93,-3.29) .. controls (6.95,-1.4) and (3.31,-0.3) .. (0,0) .. controls (3.31,0.3) and (6.95,1.4) .. (10.93,3.29)   ;
\draw [shift={(392,80)}, rotate = 88.53] [color={rgb, 255:red, 208; green, 2; blue, 27 }  ,draw opacity=1 ][line width=0.75]    (10.93,-3.29) .. controls (6.95,-1.4) and (3.31,-0.3) .. (0,0) .. controls (3.31,0.3) and (6.95,1.4) .. (10.93,3.29)   ;
\draw    (260,165) -- (344,165) ;
\draw [shift={(344,165)}, rotate = 180] [color={rgb, 255:red, 0; green, 0; blue, 0 }  ][line width=0.75]    (10.93,-3.29) .. controls (6.95,-1.4) and (3.31,-0.3) .. (0,0) .. controls (3.31,0.3) and (6.95,1.4) .. (10.93,3.29)   ;
\draw    (116,99.65) -- (150,148.95) ;
\draw [shift={(150,148.95)}, rotate = 240] [color={rgb, 255:red, 0; green, 0; blue, 0 }  ][line width=0.75]    (10.93,-3.29) .. controls (6.95,-1.4) and (3.31,-0.3) .. (0,0) .. controls (3.31,0.3) and (6.95,1.4) .. (10.93,3.29)   ;
\draw    (440,50) -- (518,50) ;
\draw [shift={(518,50)}, rotate = 176.75] [color={rgb, 255:red, 0; green, 0; blue, 0 }  ][line width=0.75]    (10.93,-3.29) .. controls (6.95,-1.4) and (3.31,-0.3) .. (0,0) .. controls (3.31,0.3) and (6.95,1.4) .. (10.93,3.29)   ;
\draw    (260,50) -- (344,50) ;
\draw [shift={(344,50)}, rotate = 179.71] [color={rgb, 255:red, 0; green, 0; blue, 0 }  ][line width=0.75]    (10.93,-3.29) .. controls (6.95,-1.4) and (3.31,-0.3) .. (0,0) .. controls (3.31,0.3) and (6.95,1.4) .. (10.93,3.29)   ;
\draw    (116,99.65) -- (150,30) ;
\draw [shift={(150,30)}, rotate = 115] [color={rgb, 255:red, 0; green, 0; blue, 0 }  ][line width=0.75]    (10.93,-3.29) .. controls (6.95,-1.4) and (3.31,-0.3) .. (0,0) .. controls (3.31,0.3) and (6.95,1.4) .. (10.93,3.29)   ;

\end{tikzpicture}

\caption{\textbf{Validation} The initial classifier is retrained in two different ways, one by a problem-space evasion attack and another one by a feature-space evasion attack, thus creating two retrained classifiers. The performance of the two classifiers is evaluated, both on problem-space evasion attack and on clean data.}
\label{validataion_pic}
\end{figure*}

A feature-space evasion attack is used for the evaluation of \emph{validation}, which is termed $\tilde{O}(h;\mathsf{D})$. This attack is tested to check whether it can be a proxy for a problem-space attack, $O(h;\mathsf{D})$. Using the retraining process mentioned above, a defense against $\tilde{O}$ is constructed; the resulting hardened classifier is termed $\tilde{h}$.
Parallelly, a \emph{baseline} $h^*$ is created. This is a robust classifier used against the target problem-space evasion attack $O$, by means of the same retraining technique, but in this case against the problem-space evasion attack $O$. Next, the performance of $\tilde{h}$ and $h^*$ against $O$ is compared.
If $\tilde{h}$ is found to be ineffective against $O$, then $\tilde{O}$ is a poor attack proxy. However, if $\tilde{h}$ is found to be robust to the problem-space attack $O$,  $\tilde{O}$ is a sufficient proxy for the target evasion attack. A graphic description of the validation process can be found in Fig.~\ref{validataion_pic} for better clearance.
Evaluation of the validation aspect is discussed in Section~\ref{S:validation}.

\begin{figure*}[h]
\tikzset{every picture/.style={line width=0.75pt}} 
\begin{tikzpicture}[x=0.75pt,y=0.75pt,yscale=-1,xscale=1]

\draw    (18,91) -- (142,91) -- (142,123) -- (18,123) -- cycle  ;
\draw (7,93) node [anchor=north west][inner sep=0.75pt]   [align=left] {\begin{minipage}[lt]{107.03pt}\setlength\topsep{0pt}
\begin{center}
\textbf{Train initial classifier}\\\textbf{h on clean data}
\end{center}

\end{minipage}};
\draw    (181,91) -- (325,91) -- (325,123) -- (181,123) -- cycle  ;
\draw (181,92) node [anchor=north west][inner sep=0.75pt]   [align=left]
{\begin{minipage}[lt]{107.03pt}\setlength\topsep{0pt}
\begin{center}
{\textbf{Obtain $\tilde{h}$ \ by retraining}\\\textbf{h with an evasion attack}}
\end{center}

\end{minipage}};
\draw    (359,91) -- (490,91) -- (490,123) -- (359,123) -- cycle  ;
\draw (350,92) node [anchor=north west][inner sep=0.75pt]   [align=left]
{\begin{minipage}[lt]{107.03pt}\setlength\topsep{0pt}
\begin{center}
{\textbf{Evaluate $\tilde{h}$ with problem-space attacks}}
\end{center}
\end{minipage}};

\draw    (142,105) -- (180,105) ;
\draw [shift={(180,105)}, rotate = 180] [color={rgb, 255:red, 0; green, 0; blue, 0 }  ][line width=0.75]    (10.93,-3.29) .. controls (6.95,-1.4) and (3.31,-0.3) .. (0,0) .. controls (3.31,0.3) and (6.95,1.4) .. (10.93,3.29)   ;
\draw    (325,105) -- (358,105) ;
\draw [shift={(358,105)}, rotate = 180] [color={rgb, 255:red, 0; green, 0; blue, 0 }  ][line width=0.75]    (10.93,-3.29) .. controls (6.95,-1.4) and (3.31,-0.3) .. (0,0) .. controls (3.31,0.3) and (6.95,1.4) .. (10.93,3.29)   ;

\end{tikzpicture}
\caption{\textbf{Generalizability} The initial classifier is retrained with an evasion attack (problem-space or feature-space), thus creating a retrained classifier. The performance of the retrained classifier is evaluated with a set of problem-space evasion attacks.} 
\label{general_pic}
\end{figure*}
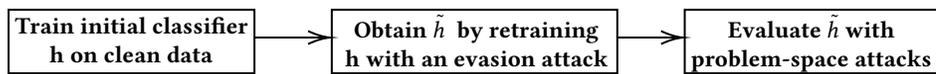

The \emph{generalizability} is evaluated slightly different from the \emph{validation} evaluation.
Again, $\tilde{O}$, a proxy attack, is considered for a problem-space evasion attack. However, in this case, it can be a feature-space or a problem-space attack. A defense $\tilde{h}$ against $\tilde{O}$ is constructed using the same process as before.
For this evaluation, instead of utilizing one evasion attack $O$, a set of target attacks $\{O_i\}$ is tested against $\tilde{h}$.
A proxy attack $\tilde{O}$ is generalizable if $\tilde{h}$ remains robust to at least most of the set of attacks $\{O_i\}$; If not, $\tilde{O}$ is declared non-generalizable.
Generalizability is the main topic of Section~\ref{PSR}.
A graphic description of the evaluation of the generalizability aspect can be found in Fig.~\ref{general_pic} for better clearance.

\section{Background}
\label{background}
\subsection{APK Structure}
\label{apk_st}
The Android PacKage (APK) is a popular format used to encapsulate and distribute Android OS's applications. It is in fact a compressed file containing the following main files/directories: \emph{manifest}, \emph{classes.dex}, \emph{layout files}, \emph{res}, and \emph{assets}.
The manifest file contains vital information to run the APK, such as the permissions that must be granted to run the application. The binary code is encapsulated in classes.dex, which can be translated to various reverse engineering languages (e.g., Smali). The layout files define how the graphic components are ordered on each page of the application (i.e., activity). The res and assets directories include resources and assets which are not code files, such as pictures and videos. A more detailed explanation can be found in~\cite{struct_dev}.

\subsection{Feature Sets and Targeted Classifiers}
\label{targets}
There are various detection systems for Android malware. Two well-known detection systems, Drebin~\cite{arp2014drebin} and MaMaDroid~\cite{onwuzurike2019mamadroidold}\footnote{These renowned detection systems have played a pivotal role in recent work on Android malware detection~\cite{demontis2017yes,chen2018android,berger2022mamadroid20,berger2021crystal,berger2020evasion,pierazzi2019intriguing,li2020adversarial,pendlebury2018tesseract,li2019android,daoudideep,kim2022mapas,gu2021graphevolvedroid,abusnaina2021dl,bromberg2020droidautoml}. 
}, are mainly distinguished by their feature sets, since their underlining learning models are standard (even though they are different models). 
The feature sets represent different aspects of Android applications.
Drebin analyzes static content from the application and creates a feature vector of the absence/presence of specific components/commands from the app. Therefore, the set of derived features of Drebin is binary. MaMaDroid creates a control flow graph of the app, then analyzes it using Hidden-Markov chains to obtain the transition probability between families/packages inside the app. Consequently, the underlining feature set of MaMaDroid holds real values. The feature sets of these two detection machines were used in this study.

\section{Experimental Methodology} 
\label{S:exp}
This section describes the experimental methodology of this work\footnote{An implementation of this work is available at Github~\cite{framework_implementation}. For an access, please contact the authors.}.
First, the evasion attacks (both problem-space and feature-space) assessed in this study are discussed (Sections~\ref{p_attacks}-\ref{f_attacks}). Then, the datasets are presented (Section~\ref{datasets}). At last, the evaluation metrics used in the experiments are reported (Sections~\ref{eval_met}).

\subsection{Problem-Space Evasion Attacks}
\label{p_attacks}
This section describes the problem-space attacks used in iterative retraining. Most of the evasion attacks in the literature are feature-space attacks. It is more complicated to create a fully functional evasion attack~
\cite{berger2020evasion,oakland2014,ndss2016,tong2019improving}. In this study, it is assumed that the attacker has no access to any information except for its malicious samples and the feature set of the targeted system. Other studies have different assumptions. For example, the attacker model proposed by Pierazzi et al.~\cite{pierazzi2020problemspace} incorporates an attacker with \textit{perfect knowledge}. In other words, the attacker runs a white-box attack and has access to the training data, the feature set, the learning algorithm, and the loss function.
In reality, most simple attackers do not have this kind of access. This current study uses \textit{zero-knowledge} attack models, with the addition of the knowledge of the feature set. Consequently, several recent problem-space evasion attacks (e.g.~\cite{pierazzi2020problemspace}) are not examined in this study, as they define different attacker models. Additional attacks that add non-operational noise to the APK (e.g.~\cite{chen2018android}) are not considered as well. This study explores attacks that modify the code of the application. Adding no-ops is a different approach, which is more recognizable with static analysis of the application (as argued in~\cite{berger2021crystal}). Therefore, this kind of evasion attack was not examined as well. 

As the target classifiers of this study analyze different types of features, each target classifier is retrained with other problem-space evasion attacks. First, four evasion attacks against Drebin and Drebin-DNN are described, followed by the evasion attacks against MaMaDroid. Drebin analyzes both the Smali code files and the manifest file (in comparison, MaMaDroid only analyzes the Smali code files). Thus, Drebin and its deep learning version, Drebin-DNN, create an extended environment for attacks. 
\subsubsection{DroidChemeleon (DC)~\cite{ndss2016}} 
The first problem-space evasion attack in this study is DroidChemeleon, a framework for evasion attacks, with various
manipulation techniques. 
The attack that was used in this work is data encryption, which targets classifiers that statically analyze the Smali code files. This attack encrypts only constant strings that are found in the Smali code. This attack is used against the Drebin and Drebin-DNN classifiers. 

\subsubsection{Random SB}
Given the feature set of Drebin, this attack enumerates the key features of an application and conceals the appearances of multiple suspicious and restricted API calls, strings, and IP addresses of the application and stores their information in the manifest file. A random subset of items from the list of key features is modified to add randomness to the attack. These items are enumerated based on the feature set of the classifier. In this way, the classifier does not recognize the appearance of these items. For example, the attack replaces an occurrence of the suspicious API call \textsf{sendTextMessage()} with an encoded version of it (\textsf{c2VuZFRleHRNZXNzYWdlKCk=}). It adds a reflection object to the Smali code that decrypts the encrypted content and runs the decrypted API call in runtime.  This attack is used against the Drebin and Drebin-DNN classifiers. 

\subsubsection{
Manifest Based attack (MB)~\cite{berger2021crystal}}In this attack, the attacker blindly changes all of the uses-permission tags in the manifest file to uses-permission-SDK-23 tags. In other words, it hides the permission requests of the app. As proven in~\cite{berger2021crystal}, the permission requests receive high weights in Drebin and, therefore, are a target for evasion attacks.
This attack is used against the Drebin and Drebin-DNN classifiers.
\subsubsection{Random MB}
A random version of the MB attack was used, as well as the original MB attack. In this version, a random subset of the permission requests of each app is concealed in a way that is similar to the original MB attack. This attack is used against the Drebin and Drebin-DNN classifiers.

\subsubsection{Structure Break (STB) Evasion Attacks~\cite{berger2022mamadroid20}}

The Structure Break evasion attacks, change the structure of an application. Move parts of the Smali code files from their original places in the application to other places and accordingly change each reference to the moved files. This attack targets Control Flow Graph (CFG) based classifiers, and specifically MaMaDroid. Two variants of the STB attack are considered~\cite{berger2022mamadroid20}, termed Random and Black-Hole Statistical evasion attacks. The difference between the two variants comprises their operation - randomly changing the flow of the app or statistically based on the samples the attacker obtains. These attacks are used against the MaMaDroid classifier. 
\subsection{Feature-Space Evasion Attacks}
\label{f_attacks}
The aim of evasion attacks is to manipulate the obtained malicious samples to appear as benign as possible. This target can be translated into an optimization problem in several ways, thus creating feature-space attacks. In this paper, it was translated one way and solved using Coordinate Greedy (Section~\ref{cg_expla}). Another way to achieve miss-classification is by adding statistical noise. The Salt \& Pepper noise attack (Section~\ref{snp}) was examined, which operates in this manner. Coordinate Greedy and Salt \& Pepper are gradient-free. Therefore, they can be applied to both DNN models and more simple ML models. Consequently, they are excellent candidates for this study, which explores different kinds of models. Both attacks can take many unrealistic approaches, as they do not consider the actual meaning of feature values. For example, flipping the bit that correlates to a GPS sensor, does not eliminate the actual use of it in the code. Therefore, as reported in the results, feature-space attacks do not serve as proxies for real problem-space evasion attacks.     

\subsubsection{Salt \& Pepper (SP)~\cite{li2020adversarial,xu2021ofei,li2021framework,kannan2020adversarial}}
\label{snp}
In the Salt \& Pepper noise attack, noise is added to the original sample until it is miss-classified. First, a random set of elements of the original sample are selected. Then, the
values of these elements are minimized or maximized (e.g., flipping ones to zeros in the binary feature values) until a miss-classification is achieved. This attack was used against all models.

\subsubsection{Coordinate Greedy (CG)~\cite{Hoos04,li2016,tong2019improving}}
\label{cg_expla}
A multi-objective optimization problem in the feature space formulates the second feature-space evasion attack. This attack is known as Coordinate Greedy:
\begin{equation}
\begin{aligned}
& \underset{x}{\text{minimize}}
& & Q(x') = f(x') + \lambda c(x',x)\\
\end{aligned}
\label{evasion-model}
\end{equation}
In this optimization problem, $x'$ is a feature vector obtained by an evasion attack that runs on a malicious seed $x$. Furthermore, $f(x')$ is the score of $x'$, given by an actual classifier $h(x') = \mathrm{sgn}(f(x'))$. The cost of changing $x'$ into $x$ is defined as $c(x',x)$. Finally, $\lambda$ is a parameter that limits the transformation cost.
The cost function c is computed by the $l_2$ norm between $x'$ and $x$: $c(x',x) = \sum_i |x'_i - x_i|^2$.
Since the features of Drebin and MaMaDroid are binary or real, the $l_2$ norm  is well received as a cost function.

 Equation~\eqref{evasion-model} presents a non-convex optimization problem. A stochastic local search called \textsl{Coordinate Greedy} (alternatively known as an iterative improvement) (CG), designed for combinatorial search domains is used to solve it. In light of the fact that CG computes a local optimum, it is common to execute it multiple times with different starting points. A basic implementation of CG from Tong et al.~\cite{tong2019improving} was used against Drebin, Drebin-DNN and MaMaDroid. This version was implemented for DNN models. To address the underlining models of Drebin (an SVM model) and MaMaDroid (RF/KNN/DT models), CG was altered to fit models that output infinite score values of Drebin and finite values (between 0-1) of MaMaDroid. 

\subsection{Datasets}
\label{datasets}

The APK file collection was obtained from the AndroZoo dataset~\cite{allix2016androzoo} (specifically from the Google Play market~\cite{GooglePlay}) and the Drebin data set~\cite{arp2014drebin}. The benign APKs were gathered from Androzoo, and the malicious ones\footnote{Recent works~\cite{berger2020evasion,wang2021advandmal,labaca2021universal,sasidharan2021prodroid,li2020adversarial} used a similar approach, choosing the Drebin dataset as their source for malicious APKs and Androzoo as their source of benign apps.} from Drebin. This dataset includes $\sim$75,000 benign apps, and $\sim$5,700 malicious apps. This ratio is according to a recent well-known work on benign/malicious populations of Android applications~\cite{pendlebury2018tesseract}.

From this dataset, $\sim$4,300 malicious and $\sim$60,000 benign APKs were used as training data, and another $\sim$1,200 malicious and $\sim$15,000 benign files as clean test data, which is another term for non-adversarial data. From the malicious training data, 40 seeds were selected as the retraining seeds and 100 seeds from the test data were chosen as the test seeds. Only 40 malicious seeds were used in the retraining process
to make the experiment more realistic. This is because in reality a small amount of malicious data is obtained and adapted by the classifier. 
This set of 40 samples was sufficient to make the basic models robust against evasion attacks. In the evaluation phase, 100 malicious seeds were used. Each seed in the retraining and evaluation phases was manipulated by a problem-space evasion attack.

The retraining and evaluation experiments were run on a Linux server (Intel(R) Xeon(R) Silver 4214R CPU @ 2.40GHz, 48 cores, and 128 GB memory, running Ubuntu 18.04).  

\subsection{Evaluation Metrics}
\label{eval_met}
Two evaluation metrics were used in this work: evasion robustness and traditional evaluation (as in~\cite{tobiyama2016malware,tong2019improving,saxe2015deep,feizollah2017androdialysis}).
The evaluation of evasion robustness was performed using 100 malicious APK seeds that were transformed using problem-space evasion attacks. Evasion robustness of 0\% means that the detection system did not identify any manipulated sample, while the evasion robustness of 100\% means that the classifier detected each manipulated sample.
The traditional evaluation metric was used as well, with clean test data, both of malicious APK and benign APK origin.
On these data, the ROC (receiver operating characteristic) curve and the corresponding AUC (area under the curve) are computed. 

\newcommand{\EVRobD}[7]{
\begin{tikzpicture}
        \begin{axis}[
        ybar=-1cm,
        axis x line*=bottom,
        axis y line*=left,
        height=8cm, width=6cm,
        bar width=1cm,
        ylabel={Evasion Robustness},
        ymax=1,
        ymin=0,
        symbolic x coords={Orig, {#1}, {#2}, {#3}},
        nodes near coords,
        nodes near coords align={vertical}          
        ]
        \addplot[red,fill] coordinates {(Orig,{#4})};
        \addplot[green,fill] coordinates {({#1},{#5})};
        \addplot[blue,fill] coordinates {({#2},{#6})};
        \addplot[black,fill] coordinates {({#3},{#7})};
                   
        \end{axis}
        
        \end{tikzpicture} 
}

\newcommand{\EVRobDnew}[7]{
\begin{subfigure}[t]{0.4\linewidth}
\centering
\begin{tikzpicture}
        \begin{axis}[
        ybar=-1cm,
        axis x line*=bottom,
        axis y line*=left,
        height=8cm, width=6cm,
        bar width=1cm,
        ylabel={Evasion Robustness},
        ymax=1,
        ymin=0,
        symbolic x coords={Orig, {#1}, {#2}, {#3}},
        nodes near coords,
        nodes near coords align={vertical}          
        ]
        \addplot[red,fill] coordinates {(Orig,{#4})};
        \addplot[green,fill] coordinates {({#1},{#5})};
        \addplot[blue,fill] coordinates {({#2},{#6})};
        \addplot[black,fill] coordinates {({#3},{#7})};
                   
        \end{axis}
        
        \end{tikzpicture}
        \caption{}
\end{subfigure}
}

\newcommand{\EVRobMnew}[4]{
\begin{subfigure}[t]{0.4\linewidth}
\centering
\begin{tikzpicture}
        \begin{axis}[
        ybar=-1cm,
        axis x line*=bottom,
        axis y line*=left,
        height=8cm, width=6cm,
        bar width=1cm,
        ylabel={Evasion Robustness},
        ymax=1,
        ymin=0,
        symbolic x coords={Orig, #4, CG(F)},
        nodes near coords,
        nodes near coords align={vertical}          
        ]
        \addplot[red,fill] coordinates {(Orig,{#1})};
        \addplot[green,fill] coordinates {({#4},{#2})};
        \addplot[blue,fill] coordinates {(CG(F),{#3})};

        \end{axis}
        \end{tikzpicture}
        \caption{}
\end{subfigure}
}

\newcommand{\EVRobM}[4]{
\begin{tikzpicture}
        \begin{axis}[
        ybar=-1cm,
        axis x line*=bottom,
        axis y line*=left,
        height=8cm, width=6cm,
        bar width=1cm,
        ylabel={Evasion Robustness},
        ymax=1,
        ymin=0,
        symbolic x coords={Orig, #4, CG(F)},
        nodes near coords,
        nodes near coords align={vertical}          
        ]
        \addplot[red,fill] coordinates {(Orig,{#1})};
        \addplot[green,fill] coordinates {({#4},{#2})};
        \addplot[blue,fill] coordinates {(CG(F),{#3})};

        \end{axis}
        \end{tikzpicture} 
}

\newcommand{\RocD}[9]{
\begin{tikzpicture}
\begin{axis}[
  xlabel={FPR},
  ylabel={TPR},
  xmin=0,
  xmax=0.001,
  xtick={0,0.0002,...,0.001},
  restrict x to domain=0:0.001,
  no markers,
  every axis plot/.append style={ultra thick},
  legend pos=south east
]
\addplot [red] table {#1};
\addplot [green] table {#2};
\addplot [blue] table {#3};
\addplot [black] table {#4};
\addlegendentry{\hspace{-.1cm}Orig (AUC={#5})}
   \addlegendentry{#9 (AUC={#6})}
   \addlegendentry{CG(F) (AUC={#7})}
   \addlegendentry{PGD(F) (AUC={#8})}
\end{axis}
\end{tikzpicture}

}

\newcommand{\RocDnew}[9]{
\begin{subfigure}[t]{0.4\linewidth}
\centering
\begin{tikzpicture}
\begin{axis}[
  xlabel={FPR},
  ylabel={TPR},
  xmin=0,
  xmax=0.001,
  xtick={0,0.0002,...,0.001},
  restrict x to domain=0:0.001,
  no markers,
  every axis plot/.append style={ultra thick},
  legend pos=south east
]
\addplot [red] table {#1};
\addplot [green] table {#2};
\addplot [blue] table {#3};
\addplot [black] table {#4};
\addlegendentry{\hspace{-.1cm}Orig (AUC={#5})}
   \addlegendentry{#9 (AUC={#6})}
   \addlegendentry{CG(F) (AUC={#7})}
   \addlegendentry{SP(F) (AUC={#8})}
\end{axis}
\end{tikzpicture}
\caption{}
\end{subfigure}
}

\newcommand{\Drebtext}[1]{
\caption{Evasion robustness of the {#1} attack (a) and performance on clean data (b) of different retrained classifiers (and original classifier) for the Drebin detection system.}
}
\newcommand{\DrebDnntext}[1]{
\caption{Evasion robustness of the {#1} attack (a) and performance on clean data (b) of different retrained classifiers (and original classifier) for the Drebin-DNN detection system.}
}
\newcommand{\MaMatext}[1]{
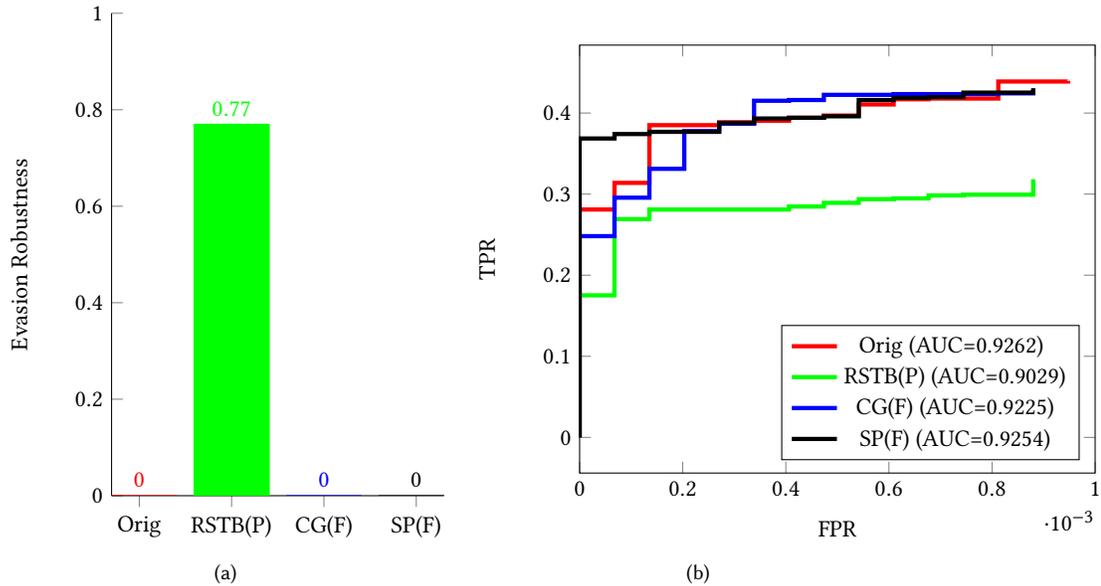
\caption{Evasion robustness of the {#1} attack (a) and performance on clean data (b) of different retrained classifiers (and original classifier) for the MaMaDroid detection system.}
}
\newcommand{\MaMaApptext}[2]{
\caption{Evasion robustness of the {#1} attack (a) and performance on clean data (b) for the MaMaDroid detection system. The underlining model is {#2}.}
}

\newcommand{\RocM}[7]{
\begin{tikzpicture}
\begin{axis}[
  xlabel={FPR},
  ylabel={TPR},
  xmin=0,
  xmax=0.001,
  xtick={0,0.0002,...,0.001},
  restrict x to domain=0:0.001,
  no markers,
  no markers,
  every axis plot/.append style={ultra thick},
  legend pos=south east
]
\addplot [red] table {#1};
\addplot [green] table {#2};
\addplot [blue] table {#3};
\addlegendentry{\hspace{-.1cm}Orig (AUC={#4})}
   \addlegendentry{{#7} (AUC={#5})}
   \addlegendentry{CG(F) (AUC={#6})}
\end{axis}
\end{tikzpicture}
}

\newcommand{\RocMnew}[7]{
\begin{subfigure}[t]{0.4\linewidth}
\centering
\begin{tikzpicture}
\begin{axis}[
  xlabel={FPR},
  ylabel={TPR},
  xmin=0,
  xmax=0.001,
  xtick={0,0.0002,...,0.001},
  restrict x to domain=0:0.001,
  no markers,
  no markers,
  every axis plot/.append style={ultra thick},
  legend pos=south east
]
\addplot [red] table {#1};
\addplot [green] table {#2};
\addplot [blue] table {#3};
\addlegendentry{\hspace{-.1cm}Orig (AUC={#4})}
   \addlegendentry{{#7} (AUC={#5})}
   \addlegendentry{CG(F) (AUC={#6})}
\end{axis}
\end{tikzpicture}
\caption{}
\end{subfigure}
}

\newcommand{\RocMnewsnp}[9]{
\begin{subfigure}[t]{0.4\linewidth}
\centering
\begin{tikzpicture}
\begin{axis}[
  xlabel={FPR},
  ylabel={TPR},
  xmin=0,
  xmax=0.001,
  xtick={0,0.0002,...,0.001},
  restrict x to domain=0:0.001,
  no markers,
  no markers,
  every axis plot/.append style={ultra thick},
  legend pos=south east
]
\addplot [red] table {#1};
\addplot [green] table {#2};
\addplot [blue] table {#3};
\addplot [black] table {#4};
\addlegendentry{\hspace{-.1cm}Orig (AUC={#5})}
   \addlegendentry{{#9} (AUC={#6})}
   \addlegendentry{CG(F) (AUC={#7})}
   \addlegendentry{SP(F) (AUC={#8})}
\end{axis}
\end{tikzpicture}
\caption{}
\end{subfigure}
}

\newcommand{\RocMwide}[7]{
\begin{tikzpicture}
\begin{axis}[
  xlabel={FPR},
  ylabel={TPR},
  xmin=0,
  no markers,
  no markers,
  every axis plot/.append style={ultra thick},
  legend pos=south east
]
\addplot [red] table {#1};
\addplot [green] table {#2};
\addplot [blue] table {#3};
\addlegendentry{\hspace{-.1cm}Orig (AUC={#4})}
   \addlegendentry{{#7} (AUC={#5})}
   \addlegendentry{CG(F) (AUC={#6})}
\end{axis}
\end{tikzpicture}
}
\newcommand{\RocMwidenew}[7]{
\begin{subfigure}[t]{0.4\linewidth}
\centering
\begin{tikzpicture}
\begin{axis}[
  xlabel={FPR},
  ylabel={TPR},
  xmin=0,
  no markers,
  no markers,
  every axis plot/.append style={ultra thick},
  legend pos=south east
]
\addplot [red] table {#1};
\addplot [green] table {#2};
\addplot [blue] table {#3};
\addlegendentry{\hspace{-.1cm}Orig (AUC={#4})}
   \addlegendentry{{#7} (AUC={#5})}
   \addlegendentry{CG(F) (AUC={#6})}
\end{axis}
\end{tikzpicture}
\caption{}
\end{subfigure}
}

\newcommand{\RocMwidenewsnp}[9]{
\begin{subfigure}[t]{0.4\linewidth}
\centering
\begin{tikzpicture}
\begin{axis}[
  xlabel={FPR},
  ylabel={TPR},
  xmin=0,
  no markers,
  no markers,
  every axis plot/.append style={ultra thick},
  legend pos=south east
]
\addplot [red] table {#1};
\addplot [green] table {#2};
\addplot [blue] table {#3};
\addplot [black] table {#4};
\addlegendentry{\hspace{-.1cm}Orig (AUC={#5})}
   \addlegendentry{{#9} (AUC={#6})}
   \addlegendentry{CG(F) (AUC={#7})}
   \addlegendentry{SP(F) (AUC={#8})}
\end{axis}
\end{tikzpicture}
\caption{}
\end{subfigure}
}

\section{Validation of Feature-Space Retraining}
\label{S:validation}

\begin{figure*}[t!]
\EVRobDnew{SB(P)}{CG(F)}{SP(F)}{0.65}{0.9}{0.8}{0.69}
\RocDnew{rocs/reports-sb-cg_clean_b_roc.txt}{rocs/reports-sb-cg_clean_p_roc.txt}{rocs/reports-sb-cg_clean_cg_roc.txt}{rocs/reports-sb-cg_clean_snp_roc.txt}{0.9973}{0.9969}{0.9967}{0.9973}{SB(P)}
\Drebtext{Random SB}
\label{eval_sb}
\end{figure*}

This paper assesses the difference in robustness of detection systems based on three different methodologies: a. simply training the system with the samples from the dataset; b. iteratively retraining the classifier using manipulated instances generated by a problem-space attack; c. iteratively retraining the classifier using manipulated instances generated by a feature-space attack. As the evaluation focuses on problem-space attacks, it is clear that the second methodology will perform best and serve as the upper bound. Thus, it is set as a baseline, to assess the degradation using either of the other two methodologies. In each evaluation, the retrained classifier on an evasion attack is signed by the name of the evasion attack (or an acronym) and the first letter of the attack type. For example, DC(P) stands for the retrained classifier on the DroidChemeleon attack, which is a problem-space evasion attack. Similarly, CG(F) refers to the retrained classifier on the CG attack, which is a feature-space evasion attack. . In order to get a clearer view of the differences between the curves, the scale of the ROC curve was changed to 0-0.001. A close examination shows that a wider scale did not depict any difference between the explored classifiers. Three ML detection systems are evaluated in this section as a proof of concept: Drebin, Drebin-DNN, and MaMaDroid. Each section describes the experiments that were carried out on the specific classifier. 

\subsection{Drebin}
\label{drebin_eval_res}
The first set of experiments was carried out on an SVM model, trained on the set of features of the well-known Android malware classifier, Drebin~\cite{arp2014drebin}.
In these experiments, the $C$ parameter was optimized using CV on clean data to the value of 1. For every experiment, the basic model was the SVM, retrained using different problem-space evasion attacks: Random SB (Section~\ref{random_sb_test}), DroidChemeleon (Section~\ref{random_sb_test}), MB (Section~\ref{droidchem_test}), and Random MB (Section~\ref{mb_test}). In these experiments, the CG and SP feature space attacks were considered, while setting $\lambda = 0.005$ for the CG attack and $\epsilon = 1000$ for the SP attack (both were the defaults of each implementation~\cite{melis2019secml,tong2019improving}). A short discussion (Section~\ref{dreb_conc}) concludes this section. 

\subsubsection{Random SB Experiment}
\label{random_sb_test}
First, the Random SB attack was run against the original Drebin classifier without any retraining process, and the classifier demonstrated an evasion robustness of 65\%.
Next, to create the baseline, Drebin was iteratively retrained with the Random SB evasion attack. As can be seen in Fig.~\ref{eval_sb}a, the retrained classifier SB(P) achieved an evasion robustness of 90\%. In comparison, the CG(F) and SP(F) retrained classifiers resulted in an evasion robustness of 80\% and 69\%, respectively.
This shows that defenses that rely on acknowledging feature-space evasion attacks do not sufficiently identify the Random SB attack. Also in this experiment, the CG(F) retrained classifier outperforms the SP(F) retrained classifier by 11\%.
Furthermore, the three retrained classifiers were found to be basically as accurate as the original Drebin classifier on clean data, with an Area under the ROC Curve (AUC) of more than 99\% (Fig.~\ref{eval_sb}b). Therefore, it is proven that a highly robust classifier for the Random SB attack (the SB(P) classifier) can be achieved without significantly harming its effectiveness on non-adversarial data.

\subsubsection{DroidChemeleon Experiment}
\label{droidchem_test}

\begin{figure*}[ht!]
\EVRobDnew{DC(P)}{CG(F)}{SP(F)}{0.8}{0.92}{0.78}{0.85}
\RocDnew{rocs/reports-droidchem-cg_clean_b_roc.txt}{rocs/reports-droidchem-cg_clean_p_roc.txt}{rocs/reports-droidchem-cg_clean_cg_roc.txt}{rocs/reports-droidchem-cg_clean_snp_roc.txt}{0.9973}{0.9965}{0.9967}{0.9973}{DC(P)}
\Drebtext{DroidChemeleon}
\label{droid_eval}
\end{figure*}

First, the DroidChemeleon attack was run on the original Drebin classifier without any retraining process, and the classifier demonstrated an evasion robustness of 80\%.
Next, Drebin was iteratively retrained with the DroidChemeleon evasion attack to create the baseline. As depicted in Fig.~\ref{droid_eval}a, the DC(P) retrained classifier achieved an evasion robustness of 92\%. In comparison, the CG(F) and SP(F) retrained classifiers resulted in an evasion robustness of 78\% and 85\%, respectively. The retraining process with the CG attack decreased the evasion robustness of the initial classifier by 2\%.
 This shows that a defense that relies on recognizing feature-space evasion attacks will not correctly identify the DroidChemeleon attack. Also in this experiment, the SP(F) retrained classifier outperformed the CG(F) retrained classifier by 8\%. 
Additionally, the three retrained classifiers were basically as accurate as the original Drebin classifier on clean data, with an AUC of more than 99\% (Fig.~\ref{droid_eval}b). Therefore, it is proven that a highly robust classifier to the DroidChemeleon attack (the DC(P) classifier) can be achieved without significant damage to its effectiveness on non-adversarial data.

\subsubsection{MB Experiment}
\label{mb_test}

\begin{figure*}[ht!]

\EVRobDnew{MB(P)}{CG(F)}{SP(F)}{0.42}{0.88}{0.53}{0.55}
\RocDnew{rocs/CG_manifest_with_manifest_retrain_clean_b_roc.txt}{rocs/CG_manifest_with_manifest_retrain_clean_p_roc.txt}{rocs/CG_manifest_with_manifest_retrain_clean_f1_roc.txt}{rocs/CG_manifest_with_manifest_retrain_clean_f2_roc.txt}{0.9973}{0.9967}{0.9973}{0.9964}{MB(P)}
\Drebtext{MB}

\label{manifest_eval}
\end{figure*}

As a first step, the MB attack was run on the original Drebin classifier without any retraining process, and the classifier demonstrated an evasion robustness of 42\%.
Then, to create the baseline, Drebin was iteratively retrained with the MB evasion attack. As illustrated in Fig.~\ref{manifest_eval}a, the MB(P) retrained classifier achieved an evasion robustness of 88\%. In comparison, the CG(F) and SP(F) retrained classifiers resulted in an evasion robustness of 53\% and 55\%, respectively. This time, the retraining processes are more distinguishable, with a wide gap of 33\% in evasion robustness, between the CG(F) and SP(F) on the one hand, and MB(P) on the other hand.
This shows that a defense that relies on recognizing feature-space evasion attacks will not identify the MB attack. In addition, the CG(F) and SP(F) classifiers achieve a similar evasion robustness. 
Lastly, the three retrained classifiers are basically as accurate as the original Drebin classifier on clean data, with an AUC of more than 99\% (Fig.~\ref{manifest_eval}b). Therefore, it is proven that a highly robust classifier for the MB attack (the MB(P) classifier) can be achieved without significant damage to its effectiveness on non-adversarial data.

\subsubsection{Random MB Experiment}
\label{random_mb_test}
\begin{figure*}[ht!]
\EVRobDnew{RMB(P)}{CG(F)}{SP(F)}{0.47}{0.84}{0.59}{0.53}
\RocDnew{rocs/CG_manifest_arbitary_clean_b_roc.txt}{rocs/CG_manifest_arbitary_clean_p_roc.txt}{rocs/CG_manifest_arbitary_clean_f1_roc.txt}{rocs/CG_manifest_arbitary_clean_f2_roc.txt}{0.9973}{0.9967}{0.9973}{0.9968}{RMB(P)}
\Drebtext{Random MB}

\label{manifest_ar_eval}
\end{figure*}

In the beginning, the Random MB attack was run on the original Drebin classifier without any retraining process, and the classifier resulted in an evasion robustness of 47\%.
Then, to create the baseline, Drebin was iteratively retrained with the Random MB evasion attack. As can be seen in Fig.~\ref{manifest_ar_eval}a, the RMB(P) retrained classifier achieved an evasion robustness of 84\%. In comparison, the CG(F) and SP(F) retrained classifiers resulted in an evasion robustness of 59\% and 53\%, respectively. As in the MB experiment, the retraining processes' are distinguishable, with a wide gap of 25\% in the evasion robustness, between the CG(F) and SP(F) on the one hand, and the RMB(P) on the other hand.
This shows that a defense that relies on recognizing feature-space evasion attacks will not identify the Random MB attack. Also in this experiment, the CG(F) retrained classifier outperformed the SP(F) retrained classifier by 6\%. 
Lastly, the three retrained classifiers are basically as accurate as the original Drebin classifier on clean data, with an AUC of more than 99\% (Fig.~\ref{manifest_ar_eval}b). Therefore, it has been proven that a highly robust classifier to the Random MB attack (the RMB(P) classifier) can be achieved without significant damage to its effectiveness on clean data.

\subsubsection{Discussion}
\label{dreb_conc}
The results presented in this section show that problem-space evasion attacks that target the Smali code are more identifiable by the Drebin classifier and its retrained classifiers than attacks that target the manifest file. The results with reference to the Random SB and DroidChemeleon experiments, show that the retrained classifiers on feature-space evasion attacks trailed the retrained classifier on the problem-space evasion attack by 7\%-11\%. However, the experiments of the MB and Random MB attacks proved that problem-space evasion attacks that target permission requests are less predictable by the retrained classifiers on feature-space evasion attacks. In these experiments, the CG(F) and SP(F) classifiers trailed the MB(P) and RMB(P) classifiers by more than 30\%. This also supports the results of previous work~\cite{berger2021crystal,berger2020evasion}. Nonetheless, these results ascertained that the same phenomena occur for classifiers that are iteratively retrained and that the difference between retraining processes of feature-space and problem-space evasion attacks is distinguishable. The next section explores the DNN version of Drebin.
\subsection{Drebin-DNN}
\label{drebin_dnn_eval_res}
In this set of experiments, a DNN model, termed Drebin-DNN, was trained on the feature set of Drebin (as in~\cite{li2020adversarial}). Drebin-DNN consisted of two fully-connected hidden layers, where each comprised 160 neurons. The activation function was ReLU. An Adam optimizer was used for optimization, with
150 epochs, and a mini-batch of size 128. The learning rate was 0.001. 

For each experiment, the basic DNN model was retrained with different problem-space evasion attacks: Random SB (Section~\ref{random_dnn_sb_test}), DroidChemeleon (Section~\ref{droidchem_dnn_test}), MB (Section~\ref{mb_dnn_test}), and Random MB (Section~\ref{random_mb_dnn_test}). The CG(F) and SP(F) feature space attacks were considered in these experiments, using the same configurations as with Drebin. A short discussion (Section~\ref{dreb_dnn_conc}) is provided at the end of this section.

\subsubsection{Random SB Experiment}
\label{random_dnn_sb_test}
\begin{figure*}[t!]
\EVRobDnew{SB(P)}{CG(F)}{SP(F)}{0.77}{0.8}{0.74}{0.78}
\RocDnew{rocs/sb_dnn_sb_b.txt}{rocs/sb_dnn_sb_p.txt}{rocs/sb_dnn_sb_cg.txt}{rocs/sb_dnn_sb_pgd.txt}{0.9964}{0.9951}{0.9936}{0.9961}{SB(P)}
\DrebDnntext{Random SB}
\label{eval_dnn_sb}
\end{figure*}

First, the Random SB attack was run against the original Drebin-DNN classifier without any retraining process, and the classifier demonstrated an evasion robustness of 77\%.
Next, to establish the baseline classifier, Drebin-DNN was iteratively retrained with the Random SB evasion attack. As demonstrated in Fig.~\ref{eval_dnn_sb}a, the SB(P) retrained classifier achieved an evasion robustness of 80\%. In comparison, the CG(F) and SP(F) retrained classifiers resulted in an evasion robustness of 74\% and 78\%, respectively.
These results show that defense that relies on recognizing feature-space evasion attacks will not correctly identify the Random SB attack. Furthermore, the retraining process with the CG attack decreased the evasion robustness of the basic classifier by 3\%.
Furthermore, the three retrained classifiers were found to be basically as accurate as the original Drebin-DNN classifier on clean data, with an AUC of more than 99\% (Fig.~\ref{eval_dnn_sb}b). This experiment shows that sometimes the retraining process in problem-space attacks does not result in a classifier that is significantly more robust.

\subsubsection{DroidChemeleon Experiment}
\label{droidchem_dnn_test}

\begin{figure*}[ht!]
\EVRobDnew{DC(P)}{CG(F)}{SP(F)}{0.84}{0.84}{0.84}{0.83}
\RocDnew{rocs/droid_dnn_droid_b.txt}{rocs/droid_dnn_droid_p.txt}{rocs/droid_dnn_droid_cg.txt}{rocs/droid_dnn_droid_pgd.txt}{0.9964}{0.9898}{0.9936}{0.9961}{DC(P)}
\DrebDnntext{DroidChemeleon}
\label{droid_dnn_eval}
\end{figure*}

First, the DroidChemeleon attack was run on the original Drebin-DNN classifier without any retraining process, and the classifier resulted in an evasion robustness of 84\%.
Next, to establish the baseline classifier, Drebin-DNN was iteratively retrained with the DroidChemeleon evasion attack. As depicted in Fig.~\ref{droid_dnn_eval}a, the DC(P) retrained classifier achieved an evasion robustness of 84\%, showing that the retraining process in this experiment did not produce any gain in evasion robustness. The CG(F) and SP(F) retrained classifiers resulted in an evasion robustness of 84\% and 83\%, respectively. This experiment also shows that sometimes the retraining process in problem-space attacks does not result in a classifier that is significantly more robust.
In addition, the three retrained classifiers are basically as accurate as the original Drebin classifier on clean data, with an AUC of $\sim 99\%$ (Fig.~\ref{droid_eval}b).

\subsubsection{MB Experiment}
\label{mb_dnn_test}

\begin{figure*}[ht!]

\EVRobDnew{MB(P)}{CG(F)}{SP(F)}{0.61}{0.88}{0.5}{0.62}
\RocDnew{rocs/mb_dnn_mb_b.txt}{rocs/mb_dnn_mb_p.txt}{rocs/mb_dnn_mb_cg.txt}{rocs/mb_dnn_mb_pgd.txt}{0.9964}{0.9968}{0.9936}{0.9961}{MB(P)}
\DrebDnntext{MB}

\label{manifest_dnn_eval}
\end{figure*}

Initialy, the MB attack was run with the original Drebin-DNN classifier without any retraining process, and the classifier demonstrated an evasion robustness of 61\%.
Then, to create the baseline, Drebin-DNN was iteratively retrained with the MB evasion attack. As illustrated in Fig.~\ref{manifest_dnn_eval}a, the MB(P) retrained classifier achieved an evasion robustness of 88\%. In comparison, the CG(F) and SP(F) retrained classifiers resulted in an evasion robustness of 50\% and 62\%, respectively. This time, the retraining processes were more distinguishable, with a wide gap of over 26\% in the evasion robustness, between the CG(F) and SP(F), on the one hand, and the MB(P) on the other. Also, the retrained CG(F) classifier decreased the evasion robustness of the original classifier by 11\%.
This shows that a defense that relies on recognizing feature-space evasion attacks will not identify the MB attack. Also, the retrained classifiers in feature-space attacks performed slightly different than in the previous experiments. In this experiment, the SP(F) retrained classifier outperformed the CG(F) retrained classifier by 12\%. 
Lastly, the three retrained classifiers are basically as accurate as the original Drebin-DNN classifier on clean data, with an AUC of more than 99\% (Fig.~\ref{manifest_dnn_eval}b). Therefore, the results demonstrate a highly robust classifier for the MB attack (the MB(P) classifier) can be achieved without significant damage to its effectiveness on non-adversarial data.

\subsubsection{Random MB Experiment}
\label{random_mb_dnn_test}
\begin{figure*}[ht!]
\EVRobDnew{RMB(P)}{CG(F)}{SP(F)}{0.59}{0.81}{0.51}{0.63}
\RocDnew{rocs/reports-manifest_arb-cg_clean_b_roc.txt}{rocs/reports-manifest_arb-cg_clean_p_roc.txt}{rocs/reports-manifest_arb-cg_clean_cg_roc.txt}{rocs/reports-manifest_arb-cg_clean_snp_roc.txt}{0.9973}{0.9969}{0.9967}{0.9973}{RMB(P)}
\DrebDnntext{Random MB}

\label{manifest_ar_dnn_eval}
\end{figure*}

First, the Random MB attack was run on the original Drebin-DNN classifier without any retraining process, and the classifier resulted in an evasion robustness of 59\%.
Then, to create the baseline, Drebin was iteratively retrained with the Random MB evasion attack. As illustrated in Fig.~\ref{manifest_ar_dnn_eval}a, the RMB(P) retrained classifier achieved an evasion robustness of 81\%. In comparison, the CG(F) and SP(F) retrained classifiers resulted in an evasion robustness of 51\% and 63\%, respectively. As in the MB experiment, the retraining processes' were  distinguishable, with a gap of 17\% in evasion robustness between the CG(F) and SP(F) on the one hand, and the RMB(P) classifier on the other hand. Also, the retrained CG(F) classifier decreased the evasion robustness of the original classifier by 8\%.
These results show that a defense that relies on recognizing feature-space evasion attacks will not identify the Random MB attack. Also, the SP(F) retrained classifier outperformed the CG(F) retrained classifier by 12\%. 
Lastly, the three retrained classifiers were basically as accurate as the original Drebin classifier on clean data, with an AUC of more than 99\% (Fig.~\ref{manifest_ar_dnn_eval}b). Therefore, the results prove that a highly robust classifier for the Random MB attack (the RMB(P) classifier) can be achieved without significant damage to its effectiveness on clean data.

\subsubsection{Discussion}
\label{dreb_dnn_conc}
In conclusion, problem-space evasion attacks that target the Smali code are more identifiable by the Drebin-DNN classifier and its retrained classifiers, as depicted in the results of the Random SB attack experiment. Also, the DroidChemeleon experiment demonstrated that retraining does not always add robustness. However, the experiments with the MB and Random MB attacks proved that problem-space evasion attacks that target permission requests are less predictable by the retrained classifiers in feature-space evasion attacks. In these experiments, the CG(F) and SP(F) classifiers trailed the MB(P) and RMB(P) classifiers by at least 12\%. As with the Drebin experiments these results support previous work~\cite{berger2021crystal,berger2020evasion}. However, these results added the understanding that the same phenomena occur for a DNN model that is iteratively retrained and that the difference between retraining processes of feature-space and problem-space evasion attacks is distinguishable as well.

Moreover, sometimes the feature-space evasion attack decreases the evasion robustness compared to the original classifier. For example, the retrained CG(F) classifier decreased the evasion robustness of the original classifier by 11\% when tested with the MB attack, and by 8\% when tested with the Random MB attack. This decrease may seem small. Nonetheless, every malicious sample that is not recognized as such damages the credibility of a malware detection system. Consequently, it is an important anecdote. Also, it is important to acknowledge that retraining with feature-space attacks does not always make the classifier more robust. This robustification process may create a bias toward unrealistic data and, as a result, affect the identification of real problem-space evasion attacks.
\subsection{MaMaDroid}
\label{mamadroid_eval_res}

The third set of experiments utilizes the feature set of MaMaDroid~\cite{onwuzurike2019mamadroid}, a well-known Android malware classifier that analyzes the control flow graph of the application. This set of experiments is different from previous experiments, as this set includes multiple models - 1NN, 3NN, and RF (as in the original MaMaDroid paper~\cite{onwuzurike2019mamadroid}). An additional DT model was examined as well, as it improves the overall evasion robustness of MaMaDroid, as proven by Berger et al.~\cite{berger2022mamadroid20}. As the four models showed similar results, three out of the four models' results are presented in Appendix~\ref{mama_additional}. The  results presented relate to the RF model, which was chosen by~\cite{onwuzurike2019mamadroid} as the most promising model. This RF model is termed MaMaDroid until the end of this study. MaMaDroid was assessed using two experiments, where each one utilized a different variant of the STB attack~\cite{berger2022mamadroid20}: Random (Section~\ref{random_stb_test}) and Black-Hole Statistical STB (Section~\ref{black_stb_test}) attacks. The CG and SP were the feature-space attacks that were examined.
A slight modification was made to the CG, since the original CG targets binary features, and MaMaDroid analyzes real-value features. Therefore, instead of flipping from 0 to 1 and vice versa, a constant value was added in a cyclic way. The value chosen was 0.01. 
The results of the Random STB experiment are described in Section~\ref{random_stb_test}. 
This section also concludes with a short discussion~\ref{mama_dis}. 

\subsubsection{Random STB Experiment}
\label{random_stb_test}

\begin{figure*}[t!]
\EVRobDnew{RSTB(P)}{CG(F)}{SP(F)}{0}{0.77}{0}{0}
\RocMnewsnp{rocs_mamadroid/cg_random_rf_clean_b_roc.txt}{rocs_mamadroid/cg_random_rf_clean_p_roc.txt}{rocs_mamadroid/cg_random_rf_clean_cg_roc.txt}{rocs_mamadroid/cg_random_rf_clean_snp_roc.txt}{0.9262}{0.9029}{0.9225}{0.9254}{RSTB(P)}

\MaMatext{Random STB}
\label{eval_stb_random_rf}
\end{figure*}

First, the Random STB attack was run on the original MaMaDroid classifier without any retraining process, and the classifier resulted in an evasion robustness of 0\%.
Then, to create the baseline, MaMaDroid was iteratively retrained with the Random STB evasion attack. As demonstrated in Fig.~\ref{eval_stb_random_rf}a, the RSTB(P) retrained classifier achieved an evasion robustness of 77\%. In comparison, the CG(F) and SP(F) retrained classifiers resulted in an evasion robustness of 0\%.
This shows that a defense that relies on recognizing feature-space evasion attacks will not identify the Random STB attack. 
Lastly, the three retrained classifiers were basically as accurate as the original MaMaDroid classifier on clean data, with an AUC of more than 90\% (Fig.~\ref{eval_stb_random_rf}b). Therefore, a highly robust classifier to the Random STB attack (the RSTB(P) classifier) can be achieved without significant damage to its effectiveness on clean data.

\subsubsection{Black-Hole Statistical STB Experiment}
\label{black_stb_test}

\begin{figure*}[t!]
\EVRobDnew{BSTB(P)}{CG(F)}{SP(F)}{0}{0.96}{0}{0}
\RocMnewsnp{rocs_mamadroid/cg_black_rf_clean_b_roc.txt}{rocs_mamadroid/cg_black_rf_clean_p_roc.txt}{rocs/cg_black_rf_clean_cg_roc.txt}{rocs/cg_black_rf_clean_snp_roc.txt}{0.9262}{0.9079}{0.9225}{0.9254}{BSTB(P)}
\MaMatext{Black-Hole Statistical STB}
\label{eval_stb_black_rf}
\end{figure*}

First, the Black-Hole Statistical STB attack was carried out on the original MaMaDroid classifier without any retraining process, and the classifier resulted in an evasion robustness of 0\%.
Then, to create the baseline, MaMaDroid was iteratively retrained with the Black-Hole Statistical STB evasion attack. As depicted in Fig.~\ref{eval_stb_black_rf}a, the BSTB(P) retrained classifier achieved an evasion robustness of 96\%. In comparison, the CG(F) and SP(F) retrained classifiers resulted in an evasion robustness of 0\%.
This shows that a defense that relies on recognizing feature-space evasion attacks will not identify the Black-Hole Statistical STB attack. 
Lastly, the three retrained classifiers were basically as accurate as the original MaMaDroid classifier on clean data, with an AUC of more than 90\% (Fig.~\ref{eval_stb_black_rf}b). Therefore, a highly robust classifier for the Black-Hole Statistical STB attack (the BSTB(P) classifier) can be achieved without significant damage to its effectiveness on non-adversarial data.

\subsubsection{Discussion}
\label{mama_dis}
As was proven in both experiments, the MaMaDroid classifier is vulnerable to STB attacks. Thus, attacks that target the structure of a sample evade classification as malicious by detection machines that analyze the control flow graph. In addition to these findings, which were proven in~\cite{berger2022mamadroid20}, this was proven in the above experiments to be true for retrained classifiers as well. The retraining process on feature-space attacks does not add defense against this kind of attack. The examined feature-space attacks, CG and SP, randomly choose an index in the feature vector and change their values. The examined problem-space attacks, STB, move partial values from one feature to another. For example, the Black-Hole Statistical STB attack moves feature values to specific parts of the feature value that mainly has values of 0 or close to it. The exact amount depends on the specific application, its structure, and its complexity. Therefore, it cannot be mimicked by feature-space evasion attacks, specifically by CG or SP. It is more complex. As proven in the previous section, retraining classifiers on feature-space evasion attacks does not cover real-life evasion attacks. 
\section{Problem-Space Evasion Attacks generalizability}
\label{PSR}
Hitherto, retrained classifiers on feature-space attacks were tested to evaluate their robustness against problem-space evasion attacks. Retraining with feature-space attacks was proven in the last section to be ineffective against most of the problem-space evasion attacks. Therefore, the feature-space attacks are eliminated from this discussion. As each problem-space evasion attack differs, in means of the weak spots it targets, it is natural to
wonder whether classifiers that are robust to one problem-space evasion attack remain robust
to other problem-space evasion attacks. In this section, each retrained classifier on a problem-space attack was tested with the other problem-space evasion attacks against its basic classifier. For example, the retrained classifier for the DroidChemeleon attack was tested with the Random SB, DroidChemeleon, MB, and Random MB attacks. In other words, these experiments attempted to check whether or not problem-space evasion attacks are generalizable. In this section, the retrained classifiers are denoted only by their initials, as all of them were retrained on problem-space evasion attacks. The metric used was evasion robustness, as it is the most suitable way to measure the differences in the identification of evasion attacks. However, instead of using the actual evasion robustness of each attack, the delta from the basic classifier was used to draw the effects of each retraining process. Due to the differences between the basic classifiers, each basic classifier is described in a seperate section, as before. First, Drebin's results are described in Section~\ref{psr_drebin}. Then, Drebin-DNN's results are described in Section~\ref{psr_drebin_dnn}. Finally, the results of MaMaDroid are shown in Section~\ref{psr_mamadroid}.

\subsection{Drebin}
\label{psr_drebin}
This section describes the results of running the evasion attacks against Drebin with each of the retrained classifiers using problem-space evasion attacks. The idea behind these experiments was to assess the generalizability of the attacks. The results are described in Table~\ref{res_drebin_psr}, which show that in most cases, the retrained classifiers using a specific problem-space attack remained robust against other problem-space evasion attacks. Specifically, retrained classifiers in the Random SB and Random MB attacks were found to be the most robust. Because both of the evasion attacks have a random element, each iteration created a somewhat different manipulated version of the original sample. Therefore, the retraining process was shown to be the most efficient. 

\begin{table}
\begin{tabular}{|c|c|c|c|c|c|}
\hline
\diagbox{AT}{DEF} & Basic & RSB & DC & MB & RMB
\\
\hline
RSB&	65\%&	+25\%&	+21\%&	+15\%&	+29\%\\
\hline
DC&	80\%&	+4\%&	+10\%&	+9\%&	+11\%\\
\hline
MB&	42\%&	+33\%&	+9\%&	+46\%&	+45\%\\
\hline
RMB&	47\%&	+24\%&	0\%&	+42\%&	+41\%\\
\hline

\end{tabular}
\caption{Generalizability of problem-space evasion attacks against Drebin. The basic classifier is Drebin. The problem-space evasion attacks (AT) are Random SB (RSB), Droichemeleon (DC), MB, and Random MB (RMB). Each retrained classifier (DEF) is described by its delta from the basic classifier (Basic).}
\label{res_drebin_psr}
\end{table}

Additionally, the nature of similar problem-space evasion attacks was revealed. In other words, in each pair of evasion attacks that target the same part of the sample, one evasion attack was found to be a great candidate for retraining a classifier against the other attack. For example, the Random SB and Droichemeleon both target the Smali code. The SB classifier was found to be more robust to Droichemeleon than the basic classifier by 4\%. On the other hand, the DC classifier was found to be more robust to the SB attack than the basic classifier by 21\%. Four percent might seem negligent. However, because each manipulated sample that evades the classifier is a threat to the public, it still matters. In addition, a similar phenomenon occurs with the MB and Random MB as they share a common target - the manifest file. The MB classifier proves to be more robust than the basic classifier by 46\% against the Random MB attack. The RMB classifier outperforms the basic classifier by 45\% when tested with the MB attack. These two retrained classifiers of the four problem-space retrained classifiers  performed the best. 

Another interesting result shows that not every retraining process on a problem-space attack creates a robust classifier against any other problem-space attack. As detailed in Table~\ref{res_drebin_psr}, the DC classifier did not increase evasion robustness in comparison to the basic classifier when the tested evasion attack was RMB.

To summarize, the results of the retrained classifiers on problem-space evasion attacks show that each one of these evasion attacks was found to be generalizable. However, the results when using the MB or the Random MB evasion attacks to retrain a classifier compared to those using the other two evasion attacks were better. The findings also show that some problem-space evasion attacks do not increase the evasion robustness of the classifier against other problem-space evasion attacks. This is understandable because retraining a classifier with new data sometimes creates a tendency in the classification process toward these new data. However, when retraining and evaluating with similar problem-space evasion attacks, this phenomenon does not appear.

\subsection{Drebin-DNN}
\label{psr_drebin_dnn}

\begin{table}
\begin{tabular}{|c|c|c|c|c|c|}
\hline
\diagbox{AT}{DEF} & Basic & RSB & DC & MB & RMB
\\
\hline
RSB&	76\%&	+6\%&	+6\%&	+10\%&	+10\%\\
\hline
DC&	85\%&	0\%&	0\%&	+2\%&	+2\%\\
\hline
MB&	58\%&	-7\%&	-7\%&	+19\%&	+19\%\\
\hline
RMB&	57\%&	-5\%&	-5\%&	+22\%&	+22\%\\
\hline

\end{tabular}
\caption{Generalizability of problem-space evasion attacks against Drebin-DNN. The basic classifier is Drebin-DNN. The problem-space evasion attacks (AT) inspected are Random SB (RSB), Droichemeleon (DC), MB, and Random MB (RMB). Each retrained classifier (DEF) is described by its delta from the basic classifier (Basic).}
\label{res_drebin_dnn_psr}
\end{table}

This section describes the results of running the evasion attacks against Drebin-DNN, with each one of the retrained classifiers using problem-space evasion attacks. The results, described in Table~\ref{res_drebin_dnn_psr}, demonstrate that in most cases, the retrained classifiers using a specific problem-space attack stayed robust against other problem-space evasion attacks. Specifically, the retrained classifiers on the MB and Random MB attacks were found to be the most robust, with a delta of at least 2\% in evasion robustness (compared to the basic classifier) when tested on other attacks. 

Additionally, the nature of the similar problem-space evasion attacks was once again confirmed. In each pair of evasion attacks that target the same part of the sample, one evasion attack was found to be a reasonable candidate for retraining a classifier against the other attack. For example, the Random SB and Droichemeleon target both the Smali code. The DC classifier was found to be more robust to the SB attack than the basic classifier by 6\%. The SB classifier was found to be as robust as the basic classifier to the Droichemeleon attack. In addition, a similar phenomenon was revealed for the MB and Random MB because they share a common target, i.e., the manifest file. The MB classifier was shown to be more robust than the basic classifier by 22\% against the Random MB attack. The RMB classifier outperformed the basic classifier by 19\% when tested on the MB attack.

Another interesting set of results shows that not every retrained classifier on a problem-space attack correctly classifies each type of problem-space evasion attack. The results presented in Table~\ref{res_drebin_dnn_psr} show that the DC classifier and the SB classifier were evaded by an additional 7\% in comparison to the basic classifier when the evasion attack was MB and an additional 5\% when the evasion attack was Random MB. 

To summarize this section, the results of the retrained classifiers on problem-space evasion attacks showed that using the MB or the Random MB evasion attacks to retrain a classifier is better than using the other two evasion attacks. Each of these evasion attacks was found to be generalizable. Some problem-space evasion attacks damaged the evasion robustness of the classifier against other problem-space evasion attacks. This understandable because retraining a classifier with new data sometimes creates a tendency in the classification process toward these new data. However, when retraining and evaluating with similar problem-space evasion attacks, this phenomenon does not occur.

Two insights merged from the generalizability assessments of Drebin and Drebin-DNN. First, the attacks that are generalizable in both detection systems is the MB and RMB attack, which targets the permission requests, a component that is of a high value for both Drebin and Drebin-DNN (as was proven in~\cite{berger2021crystal}). On the other hand, a less efficient retraining process was achieved when retraining on the DroidChemeleon attack. This attack achieved the lowest evasion robustness deltas among the other retrained classifiers.

\subsection{MaMaDroid}
\label{psr_mamadroid}

\begin{table}[!]
\begin{tabular}{|c|c|c|c|c|}
\hline
\diagbox{AT}{DEF} & Basic & RSTB & BSTB 
\\
\hline
RSTB&	0\%&	+79\%&	+79\%\\
\hline
BSTB&	0\%&	+96\%&+96\%\\
\hline
\end{tabular}
\caption{Generalizability of problem-space evasion attacks against MaMaDroid. The basic classifier is MaMaDroid. The problem-space evasion attacks (AT) inspected are Random STB (RSTB) and Black-Hole Statistical STB (BSTB). Each retrained classifier (DEF) is described by its delta from the basic classifier (Basic). }
\label{res_mamadroid_psr}
\end{table}

This section describes the results of running the evasion attacks against MaMaDroid, with each one of the retrained classifiers using the STB evasion attacks. The idea behind these experiments was to explore the generalizability of these evasion attacks. As in Section~\ref{mamadroid_eval_res}, the results of the RF model are described in this section. The results of the other three models are provided in the Appendix~\ref{mamadroid_p_other}. The results of the generalizability of the evasion attacks against MaMaDroid, presented in Table~\ref{res_mamadroid_psr}, demonstrate that the retrained classifiers had the same results for each one of the evasion attacks. In other words, the evasion robustness gained from retraining with each of the STB attacks resulted in the same evasion robustness against each one of the other evasion attacks. These attacks are also generalizable. The evasion robustness of the Random STB attack was found to be 79\% more than the basic classifier. The retrained classifiers gained an evasion robustness of 96\% more than the basic classifier against the Black-Hole STB attack. 

The STB attacks share a common approach, as stated in the paper in which they originated~\cite{berger2022mamadroid20}. Therefore, it is not surprising that the classifiers that were retrained on them would achieve identical results when confronting with the same attack. Still, it is important to address this similarity after the retraining process. As the results show, for any specific retrained classifier, the difference in evasion robustness remained the same, namely - the random STB attack was more evasive than the Black-Hole Statistical STB attack.

To summarize, the results of the comparison between the retrained classifiers on problem-space evasion attacks showed that similar problem-space evasion attacks create similar effective retrained classifiers (i.e., the cases of the STB attacks). Also, a retrained classifier on one type of problem-space evasion attack can be sufficiently robust against another type of evasion attack (e.g., the MB classifier and the DC attack). However, it is not guaranteed, as exampled by the DC classifier (of the Drebin-DNN system) which decreased the initial evasion robustness against the MB and RMB attacks. Also, at times, retraining on one problem-space evasion attack will degrade the evasion robustness against other problem-space evasion attacks. This is understandable, as retraining on one attack may create a tendency in the classification process toward this attack. Consequently, the retrained classifier is less efficient against other attacks.

\section{Related Work}
\label{related-work}

This section describes some of the related work on evasion attacks and defenses in ML-based malware detection; A general insightful survey can be found in Vorobeychik and Kantarcioglu~\cite{Vorobeychik18book}. 

\subsection{Evasion attacks and Adversarial Examples} 
One of the well-celebrated problem-space evasion attacks on ML was suggested by Fogla et al.~\cite{Fogla06,Fogla06b}. This attack targeted an anomaly-based IDS. More recent work used several methods such as obfuscation, reflection, and adding noise to the original sample. Demontis et al.~\cite{demontis2017yes} used the obfuscation of suspicious constant values in the APK such as package names and API calls.  DaDidroid~\cite{ikram2019dadidroid} explored a similar approach to that of Demontis et al.~\cite{demontis2017yes} using obfuscation as well. Rastogi et al.~\cite{rastogi2013droidchameleon} presented an evasion attack, which combined the approach of Demontis et al.~\cite{demontis2017yes}  with the addition of the reflection approach (reflection loads additional code at runtime).

Another form of problem-space evasion attacks adds noises to the application to achieve miss-classification. A stub function, one of the famous types of noise, is a non-operational function, that does not change any functionality of the app. However, it slightly modifies the original order of API calls in the application. A well-known example of a stub function addition is Android HIV~\cite{chen2018android}, where the authors added suspicious functions that are not invoked to evade the Drebin classifier, and stub functions to evade the MaMaDroid classifier. A more recent example of adding noise to the app is the work of Pierazzi et al.~\cite{pierazzi2020problemspace}, where the authors implanted functions from benign apps in malicious apps to be miss-classified by the Drebin and Sec-SVM~\cite{demontis2017yes} classifiers. Rosenberg et al.~\cite{rosenberg2018generic} generated evasion attacks using three approaches, namely: the addition of non-operational functions, obfuscation of strings, and encoding of API calls. Cara et al.~\cite{cara2020feasibility} added non-invoked classes to the end of functions to achieve miss-classification. 

In addition to problem-space evasion attacks, 
a vast amount of feature-space was explored~\cite{Athalye18,Carlini17,iclr2015,pkdd2013,kdd2004,kdd2005,asiaccs2006,icml2016,Biggio13,Biggio15,icml2016,nips2014,li2016,Nelson12,Vorobeychik14}.
The most common domain of feature-space evasion attacks is image classification and, more specifically, the art of evading deep neural networks~\cite{iclr2015,iclr2016,papernot2016,CCS2016,xu2021towards,cheng2020deep}.
Several approaches attempted to generate a new version of adversarial examples in the real world. An example of one is the use of stickers on a stop sign to achieve misclassification. These attacks are considered analogous to problem-space evasion attacks~\cite{CCS2016,Evtimov18}.

\subsection{Robustification Methods Against Evasion Attacks}
The first approach to robustification of detection systems was devised by Dalvi et al.~\cite{kdd2004}. The authors formulated robust classification as a
game between the classifier and the attacker and produced an optimal
classifier strategy.
Other approaches defined robust classification as a Min-Max loss problem. In this approach, the adversary's goal is the maximization of the loss of the defender by means of small modifications of the feature values~\cite{Teo07,kdd2012,Madry18,Raghunathan18,Wong18}. Additional approaches to design robust classifiers are the use of a non-zero-sum game~\cite{Bruckner12,kdd2011,nips2014, li2016,icml2016}, or a Stackelberg game, where the learner is the leader, while the attacker is the follower~\cite{kdd2011,nips2014,icml2016,aisec2016}.
Finally, iterative retraining defense mechanisms have been proposed, both for general evasion attacks~\cite{li2016,icml2016}, and specifically for deep learning detection systems of computer vision~\cite{iclr2015,iclr2016,Madry18}.
This list of the robustification efforts shares one important characteristic: the underlining attacks that are used for the robustification of basic classifiers are feature-space attacks, which, as this paper proves, hardly constitute a proxy for practical problem-space evasion attacks.

\section{Discussion and Conclusion}
\label{conclusion}

This paper presents the examination of the robustification efforts of ML-based malware detection systems, using problem-space and feature-space evasion attacks. Several observations can be drawn from this work. First, a defense that is based on learning feature-space evasion attacks apparently fails to achieve high robustness against problem-space evasion attacks. This observation raises certain doubts about the commonplace focus on such attacks as an upgrade to ML defense and suggests that the practical usefulness of such techniques cannot be overlooked.
However, there is a difference between the types of problem evasion attacks. The attacks that change the content of the Smali code itself are more predictable and usually do not greatly change feature values. Consequently, retrained feature-space classifiers   can be used to detect such attacks. In comparison, attacks that target the manifest file change feature values that are less commonly changed. Therefore, the retrained feature-space classifiers fail to identify this kind of attack. For example, the DroidChemeleon attack changes constant string values that might not take part in the feature extraction of Drebin. On the contrary, the MB attacks change permission requests, which are correlated to features that Drebin analyzes and are highly weighted by it. Therefore, the MB attacks are harder to identify.

Second, the complexity of problem-space evasion attacks is seldomly hard to map to the feature space. Some of the changes that occur during a problem-space evasion attack, as mentioned with regards to DroidChemeleon, are not translated to a change in the feature vector extracted by the ML. Consequently, this evasion attack has high evasion robustness in reference to each of the retrained classifiers, including the retrained classifiers in feature-space attacks. However, some changes, such as the case of STB evasion attacks, move feature values inside the feature vector, that cannot be forecast and mathematically analyzed in a general way. The art of devising these attacks is not very complicated to implement. Also, these attacks do not alter the functionality of the sample. But determining which part of the feature vector moves to other parts of it is very difficult, and can take infinite time for mathematical-based approaches like retraining a classifier on feature-space attacks.

Third, some problem-space retraining can increase robustness against other problem-space evasion attacks. Two problem-space evasion attacks that target the same part of the APK sample seem to have common elements. As was proven, it is true that problem-space retraining with an evasion attack that targets the Smali code was efficient against another evasion attack that targets the Smali code. The same idea is true for attacks that target the manifest file. However, not every classifier retrained with one type of problem-space evasion attack, is robust against another type of problem-space evasion attack. These phenomena show that the robustness gained from retraining may harm the basic detection of malicious entities, as it creates a bias towards a specific kind of evasion attack, which does not reflect reality. In the daily apps markets, various kinds of malicious applications can be found, in which some encapsulate different kinds of evasion attacks. Therefore, this kind of bias harms the credibility of retraining on problem-space evasion attacks.

The gaps between retraining processes of different types of evasion attacks were drawn. However, these findings are just the tip of the iceberg. Retraining with unrealistic feature-space evasion attacks, and counting on this process as a robustness technique is not realistic.

This work used a particular class of feature-space attacks, with the $l_2$ norm to measure the cost of feature modifications, and stochastic local search to compute evasion attempts.
Future evasion attack algorithms may be developed which might be better than the evasion attacks that were used in this study.
Testing new feature-space attacks is left for future work. 

Other directions will also be explored in future work. For example, the problem-space evasion attacks that were tested targeted a specific part of the application. More complex evasion attacks that target both the Smali code files and the manifest file may create a greater challenge for Android malware detection systems. Their effects should be evaluated, and also the effect of the retraining process of basic classifiers. In addition, two types of feature vectors were tested: Drebin's and MaMaDroid's. Each feature vector depicts other aspects of the sample. Different types of feature vectors, such as features extracted from the dynamic analysis of the application, are missing from the analysis provided in this study. These feature vectors are a base for the future extension of this work.

\section*{Acknowledgments}
This work was supported by the Ariel Cyber Innovation Center in conjunction with the Israel National Cyber Directorate of the Prime Minister's Office. 

{\normalsize
\bibliographystyle{ACM-Reference-Format}
\bibliography{bibliography}}

\appendix
\section*{Appendix}
\section{MaMaDroid Models - Feature Space evaluation}
\label{mama_additional}
This appendix provides the results of the evaluation of retraining based on feature-space attacks of the three models that were inspected in MaMaDroid, other than RF: 1NN, 3NN, DT. Due to the fact that some of the models had probabilities of 100\% and 0\% between the classes, fewer points in the roc evaluation, the roc curves were enlarged to the default range of 0-1. The results of the 1NN model are presented in Section~\ref{1nn_eval}.
The results of the 3NN model are discussed in Section~\ref{3nn_eval}.
The results of the DT model are presented in Section~\ref{dt_eval}.
\subsection{1NN}
\label{1nn_eval}
This section describes the evaluation of the retrained classifiers of the 1NN model. First, the Random STB experiment is presented in Section~\ref{random_stb_test_1nn}, and then, the Black-Hole Statistical STB experiment is described in Section~\ref{black_stb_test_1nn}.
\subsubsection{Random STB Experiment}
\label{random_stb_test_1nn}

\begin{figure*}[t!]
\EVRobDnew{RSTB(P)}{CG(F)}{SP(F)}{0}{0.8}{0}{0.41}
\RocMwidenewsnp{rocs_mamadroid/cg_random_1nn_clean_b_roc.txt}{rocs_mamadroid/cg_random_1nn_clean_p_roc.txt}{rocs_mamadroid/cg_random_1nn_clean_cg_roc.txt}{rocs_mamadroid/cg_random_1nn_clean_snp_roc.txt}{0.9184}{0.9184}{0.9184}{0.9183}{RSTB(P)}

\MaMaApptext{Random STB}{1NN}
\label{eval_stb_random_1nn}
\end{figure*}

First, the Random STB attack was run on the original MaMaDroid classifier without any retraining process, and the classifier resulted in an evasion robustness of 0\%.
Then, to create the baseline, MaMaDroid was iteratively retrained with the Random STB attack. As depicted in Fig.~\ref{eval_stb_random_1nn}a, the RSTB(P) retrained classifier achieved an evasion robustness of 80\%. In comparison, the CG(F) retrained classifier resulted in an evasion robustness of 0\%, and the SP(F) retrained classifier obtained evasion robustness of 41\%. The retraining processes are fully distinguishable, with a wide gap of 39\% in evasion robustness between the RSTB(P) and the CG(F) and SP(F) classifiers.
This shows that defense that relies on recognizing feature-space evasion attacks will not fully identify the Random STB attack. 
Lastly, the three retrained classifiers were basically as accurate as the original MaMaDroid classifier on clean data, with an AUC of more than 91\% (Fig.~\ref{eval_stb_random_1nn}b). Therefore, a highly robust classifier for the Random STB attack (the RSTB(P) classifier) can be achieved without significant damage to its effectiveness on non-adversarial data.

\subsubsection{Black-Hole Statistical STB Experiment}
\label{black_stb_test_1nn}

\begin{figure*}[t!]
\EVRobDnew{BSTB(P)}{CG(F)}{SP(F)}{0}{0.96}{0}{0.41}
\RocMwidenewsnp{rocs_mamadroid/cg_black_1nn_clean_b_roc.txt}{rocs_mamadroid/cg_black_1nn_clean_p_roc.txt}{rocs_mamadroid/cg_black_1nn_clean_cg_roc.txt}{rocs_mamadroid/cg_black_1nn_clean_snp_roc.txt}{0.9184}{0.9184}{0.9184}{0.9183}{BSTB(P)}
\MaMaApptext{Black-Hole Statistical STB}{1NN}
\label{eval_stb_black_1nn}
\end{figure*}

At first, the Black-Hole statistical STB attack was run on the original MaMaDroid classifier Without any retraining process, and the classifier resulted in an evasion robustness of 0\%.
Then, to create the baseline, MaMaDroid was iteratively retrained with the Black-Hole Statistical STB attack. As illustrated in Fig.~\ref{eval_stb_black_1nn}a, the BSTB(P) retrained classifier achieved an evasion robustness of 96\%. In comparison, the CG(F) and SP(F) retrained classifiers resulted in an evasion robustness of 0\% and 41\%, respectively. The retraining processes are fully distinguishable, with a wide gap of 55\% in the evasion robustness.
This shows that a defense that relies on recognizing feature-space evasion attacks will not identify the Black-Hole Statistical STB attack. 
Lastly, the three retrained classifiers are basically as accurate as the original MaMaDroid classifier on clean data, with an AUC of more than 91\% (Fig.~\ref{eval_stb_black_1nn}b). Therefore, a highly robust classifier for the Black-Hole Statistical STB attack (the BSTB(P) classifier) can be achieved without significant damage to its effectiveness on non-adversarial data.

\subsection{3NN}
\label{3nn_eval}
This section describes the evaluation of the retrained classifiers of the 3NN model. First, the Random STB experiment is detailed in Section~\ref{random_stb_test_3nn}. Then, the Black-Hole Statistical STB experiment is described in Section~\ref{black_stb_test_3nn}.
\subsubsection{Random STB Experiment}
\label{random_stb_test_3nn}

\begin{figure*}[t!]
\EVRobDnew{RSTB(P)}{CG(F)}{SP(F)}{0}{0.75}{0}{0.42}
\RocMwidenewsnp{rocs_mamadroid/cg_random_3nn_clean_b_roc.txt}{rocs_mamadroid/cg_random_3nn_clean_p_roc.txt}{rocs_mamadroid/cg_random_3nn_clean_cg_roc.txt}{rocs_mamadroid/cg_random_3nn_clean_snp_roc.txt}{0.9336}{0.9336}{0.9336}{0.9335}{RSTB(P)}
\MaMaApptext{Random STB}{3NN}
\label{eval_stb_random_3nn}
\end{figure*}

First, the Random STB attack was run on the original MaMaDroid classifier without any retraining process, and the classifier resulted in an evasion robustness of 0\%.
Then, to create the baseline, MaMaDroid was iteratively retrained with the Random STB attack. As illustrated in Fig.~\ref{eval_stb_random_3nn}a, the RSTB(P) retrained classifier achieved an evasion robustness of 75\%. In comparison, the CG(F) and SP(F) retrained classifiers resulted in an evasion robustness of 0\% and 42\%, respectively. The retraining processes are fully distinguishable, with a wide gap of 33\% in evasion robustness.
This shows that defense that relies on recognizing feature-space evasion attacks will not identify the Random STB attack. 
Lastly, the three retrained classifiers are basically as accurate as the original MaMaDroid classifier on clean data, with an AUC of more than 93\% (Fig.~\ref{eval_stb_random_3nn}b). Therefore, a highly robust classifier for the Random STB attack (the RSTB(P) classifier) can be achieved without significant damage to its effectiveness on non-adversarial data.

\subsubsection{Black-Hole Statistical STB Experiment}
\label{black_stb_test_3nn}

\begin{figure*}[t!]
\EVRobDnew{BSTB(P)}{CG(F)}{SP(F)}{0}{0.98}{0}{0.41}
\RocMwidenewsnp{rocs_mamadroid/cg_black_3nn_clean_b_roc.txt}{rocs_mamadroid/cg_black_3nn_clean_p_roc.txt}{rocs_mamadroid/cg_black_3nn_clean_cg_roc.txt}{rocs_mamadroid/cg_black_3nn_clean_snp_roc.txt}{0.9336}{0.9336}{0.9336}{0.9335}{BSTB(P)}
\MaMaApptext{Black-Hole Statistical STB}{3NN}
\label{eval_stb_black_3nn}
\end{figure*}
At first, the Black-Hole statistical STB attack was run on the original MaMaDroid classifier Without any retraining process, and the classifier resulted in an evasion robustness of 0\%.
Then, to create the baseline, MaMaDroid was iteratively retrained with the Black-Hole Statistical STB attack. As demonstrated in Fig.~\ref{eval_stb_black_3nn}a, the BSTB(P) retrained classifier achieved an evasion robustness of 98\%. In comparison, the CG(F) and SP(F) retrained classifiers resulted in an evasion robustness of 0\% and 41\%, respectively. The retraining processes are fully distinguishable, with a wide gap of 57\% in the evasion robustness.
This shows that defense that relies on recognizing feature-space evasion attacks will not identify the Black-Hole Statistical STB attack. 
Lastly, the two retrained classifiers are basically as accurate as the original MaMaDroid classifier on clean data, with an AUC of more than 93\% (Fig.~\ref{eval_stb_black_3nn}b). Therefore, a highly robust classifier for the Black-Hole Statistical STB attack (the BSTB(P) classifier) can be achieved without significant damage to its effectiveness on non-adversarial data.

\subsection{DT}
\label{dt_eval}
This section describes the evaluation of retrained classifiers on feature-space attacks of the DT model. The DT model was suggested in~\cite{berger2022mamadroid20} as more effective than the 3 models of the original MaMaDroid~\cite{onwuzurike2019mamadroid}. Therefore, it was considered as well. First, the Random STB experiment is described in Section~\ref{random_stb_test_dt} and then, the Black-Hole Statistical STB experiment is detailed in Section~\ref{black_stb_test_dt}.
\subsubsection{Random STB Experiment}
\label{random_stb_test_dt}

\begin{figure*}[t!]
\EVRobDnew{RSTB(P)}{CG(F)}{SP(F)}{0}{0.73}{0}{0}
\RocMwidenewsnp{rocs_mamadroid/cg_random_dt_clean_b_roc.txt}{rocs_mamadroid/cg_random_dt_clean_p_roc.txt}{rocs_mamadroid/cg_random_dt_clean_cg_roc.txt}{rocs_mamadroid/cg_black_dt_clean_snp_roc.txt}{0.8995}{0.8932}{0.8956}{0.8938}{RSTB(P)}
\MaMaApptext{Random STB}{DT}
\label{eval_stb_random_dt}
\end{figure*}

First, the Random STB attack was run on the original MaMaDroid classifier without any retraining process, and the classifier resulted in an evasion robustness of 0\%.
Then, to create the baseline, MaMaDroid was iteratively retrained with the Random STB attack. As depicted in Fig.~\ref{eval_stb_random_dt}a, the RSTB(P) retrained classifier achieved an evasion robustness of 73\%. In comparison, the CG(F) and SP(F) retrained classifier resulted in an evasion robustness of 0\%. The retraining processes are fully distinguishable, with a wide gap of 73\% in evasion robustness.
This shows that defense that relies on recognizing feature-space evasion attacks will not identify the Random STB attack. 
Lastly, the two retrained classifiers are basically as accurate as the original MaMaDroid classifier on clean data, with an AUC of more than 89\% (Fig.~\ref{eval_stb_random_dt}b). Therefore, a highly robust classifier for the Random STB attack (the RSTB(P) classifier) can be achieved without significant damage to its effectiveness on non-adversarial data.

\subsubsection{Black-Hole Statistical STB Experiment}
\label{black_stb_test_dt}

\begin{figure*}[t!]
\EVRobDnew{BSTB(P)}{CG(F)}{SP(F)}{0}{0.94}{0}{0}
\RocMwidenewsnp{rocs_mamadroid/cg_black_dt_clean_b_roc.txt}{rocs_mamadroid/cg_black_dt_clean_p_roc.txt}{rocs_mamadroid/cg_black_dt_clean_cg_roc.txt}{rocs_mamadroid/cg_black_dt_clean_snp_roc.txt}{0.8995}{0.8979}{0.8956}{0.8939}{BSTB(P)}
\MaMaApptext{Black-Hole Statistical STB}{DT}
\label{eval_stb_black_dt}
\end{figure*}

At first, the Black-Hole statistical STB attack was run on the original MaMaDroid classifier Without any retraining process, and the classifier resulted in an evasion robustness of 0\%.
Then, to create the baseline, MaMaDroid was iteratively retrained with the Black-Hole Statistical STB attack. As illustrated in Fig.~\ref{eval_stb_black_dt}a, the BSTB(P) retrained classifier achieved an evasion robustness of 94\%. In comparison, the CG(F) and SP(F) retrained classifiers resulted in an evasion robustness of 0\%. The retraining processes are fully distinguishable, with a wide gap of 94\% in evasion robustness.
This shows that defense that relies on recognizing feature-space evasion attacks will not identify the Black-Hole Statistical STB attack. 
Lastly, the two retrained classifiers are basically as accurate as the original MaMaDroid classifier on clean data, with an AUC of more than 89\% (Fig.~\ref{eval_stb_black_3nn}b). Therefore, a highly robust classifier for the Black-Hole Statistical STB attack (the BSTB(P) classifier) can be achieved without significant damage to its effectiveness on non-adversarial data.

\section{MaMaDroid Models - Problem-Space evaluation}
\label{mamadroid_p_other}
This section reports the evaluation of 1NN, 3NN and DT retrained classifiers that confront the three STB attacks, to check whether or not the attacks are generalizable. This section is a continuation of Section~\ref{psr_mamadroid}, where only the RF model was described. The results show that the 1NN and 3NN retrained classifiers (Tables~\ref{res_mamadroid_psr_1nn} and \ref{res_mamadroid_psr_3nn}) show results similar to those of the retrained RF classifiers. The DT retrained classifiers (Table~\ref{res_mamadroid_psr_dt}) were slightly less efficient. However, the same behavior that was established before can be seen with these results: Each retrained classifier with an STB attack, no matter which of the attacks was used, resulted in the same evasion robustness against the STB attacks.
Therefore, as with the RF model, these STB attacks were proven to be generalizable.

\begin{table}[!]
\begin{tabular}{|c|c|c|c|c|}
\hline
\diagbox{AT}{DEF} & Basic & RSTB & BSTB 
\\
\hline
RSTB&	0\%&	+79\%&	+79\%\\
\hline
BSTB&	0\%&	+96\%&	+96\%\\
\hline

\end{tabular}
\caption{Evasion Robustness of problem-space retrained classifiers against other problem-space evasion attacks. The basic classifier is MaMaDroid's 1NN. The inspected problem-space evasion attacks (AT) are Random STB (RSTB) and Black-Hole Statistical STB (BSTB). Each retrained classifier (DEF) is described by its delta from the basic classifier (Basic). }
\label{res_mamadroid_psr_1nn}
\end{table}

\begin{table}[!]
\begin{tabular}{|c|c|c|c|c|}
\hline
\diagbox{AT}{DEF} & Basic & RSTB & BSTB 
\\
\hline
RSTB&	0\%&	+79\%&	+79\%\\
\hline
BSTB&	0\%&	+96\%&	+96\%\\
\hline

\end{tabular}
\caption{Evasion Robustness of problem-space retrained classifiers against other problem-space evasion attacks. The basic classifier is MaMaDroid's 3NN. The problem-space evasion attacks (AT) are Random STB (RSTB) and Black-Hole Statistical STB (BSTB). Each retrained classifier (DEF) is described by its delta from the basic classifier (Basic). }
\label{res_mamadroid_psr_3nn}
\end{table}

\begin{table}[!]
\begin{tabular}{|c|c|c|c|c|}
\hline
\diagbox{AT}{DEF} & Basic & RSTB & BSTB 
\\
\hline
RSTB&	0\%&	+69\%&	+69\%\\
\hline
BSTB&	0\%&	+94\%&	+94\%\\
\hline

\end{tabular}
\caption{Evasion Robustness of problem-space retrained classifiers against other problem-space evasion attacks. The basic classifier is MaMaDroid's DT. The problem-space evasion attacks (AT) are Random STB (RSTB) and Black-Hole Statistical STB (BSTB). Each retrained classifier (DEF) is described by its delta from the basic classifier (Basic). }
\label{res_mamadroid_psr_dt}
\end{table}


\end{document}